\documentclass[12pt,a4,doublespacing]{elsart}

\usepackage{natbib}

 \usepackage{epsf}

\usepackage{amssymb}
\usepackage{longtable}
\def\Vec#1{\mbox{\boldmath $#1$}}

\begin{document}

\begin{frontmatter}



\title{SIMULATION AND THEORY 
OF THE IMPACT OF TWO-DIMENSIONAL ELASTIC DISKS}


\author{Hisao Hayakawa\corauthref{hisao}}
\ead{hisao@yuragi.jinkan.kyoto-u.ac.jp}
\author{, Hiroto Kuninaka}
\corauth[hisao]{Corresponding author. Tel.: +81-75-753-6782;
 fax: +81-75-753-2931.}
\address{Graduate School of Human and Environmental Studies,
  Kyoto University, Sakyo-ku, Kyoto, 606-8501, Japan}

\begin{abstract}
The impact of a two-dimensional elastic disk with a wall is 
numerically studied. It is clarified that the coefficient of restitution (COR)
decreases with the impact velocity. The result is not consistent with 
the recent quasi-static theory of inelastic collisions even for very slow
impact. This suggests that the elastic model cannot be used in
the  quasi-static limit. 
A new quasi-static theory of impacts is
 proposed, in which the effect of thermal diffusion is dominant.
 The abrupt decrease of COR has been found
 due to the plastic deformation of the disk, 
which is assisted by the initial internal motion.
\end{abstract}


\end{frontmatter}

\section{Introduction}

\par\indent
The collision of particles with the internal degrees of freedom is
inelastic in general. 
The inelastic collisions are abundant in nature(Goldsmith, 1960).
 Examples can be seen
in collisions of atoms, molecules, elastic materials, balls
in sports, and so on.
The study of inelastic collisions 
will be able to be widely accepted as one of fundamental subjects in physics,
because they are almost always discussed 
 in textbooks of  elementary classical mechanics.

Physicists realize that inelastic collisions can be a fashionable
 subject in physics from
 recent extensive interest in granular
 materials(Kadanoff, 1999; de Gennes, 1999).
In fact, granules consist of macroscopic dissipative particles. 
Therefore, the decision of interaction among particles is obviously
 important. We believe that static interactions among granular
 particles can be described by the theory of
 elasticity(Love, 1927; Landau et al., 1960; Johnson, 1985;
 Hills et al., 1993). For example,
 the normal compression may be described by the Hertzian contact
 force(Hertz, 1882) and the shear force may be represented by the
 Mindlin force(Mindlin, 1949). The dynamical part related to the
 dissipation, however, cannot be described by any reliable physical theory.  
Thus, the distinct element method (Cundall \& Struck, 1979)
 which is one of the
 most popular models to simulate collections of granular particles
 contains some dynamical undetermined parameters.
  In other words, to determine such the parameters is
 important  for both granular physics and fundamental
 physics.

The normal impact of macroscopic materials is 
characterized by the coefficient of restitution (COR)
defined by
\begin{equation}
e=-v_{r}/v_{i} ,
\end{equation}
where $v_i$ and $v_r$ are the relative velocities of incoming and outgoing
 particles respectively.
COR $e$ had been believed to be
  a material constant, since the classical experiment by Newton (1962).
 In general, however, experiments show that COR
 for three dimensional materials 
is not a constant even in approximate sense
 but
 depends strongly on
 the impact velocity(Goldsmith, 1960; Sondergaard et al., 1990;
 Bridges et al., 1984;
 Supulver et al., 1995; Giese et al., 1996; Aspelmeier et al., 1998;
 Basile et al., 2000; Labous et al., 1995).

The origin of the dissipation in inelastic collisions is the transfer of
the kinetic energy of the center of mass 
into the internal degrees of freedom during the impacts.
Systematic theoretical investigations of the impact have begun with the paper
 by Kuwabara and Kono (1987). 
Taking into account the
viscous motion among the internal degrees of freedom, they derived 
the equation of the macroscopic deformation.
Later, Brilliantov {\it et al.} (1996) and
Morgado and Oppenheim (1997)  derived the identical equation
to eq.(\ref{oppenheim}). In
particular, the derivation by Morgado and Oppenheim (1997) 
is based on the standard technique of nonequilibrium statistical
mechanics to extract the slow mode among the fast many modes which can
be regarded as the thermal reservoir with constant temperature (see Appendix).
Furthermore, Brilliantov {\it et al.} (1996)
compared their theoretical results with experimental results.
Thus, the quasi-static theory has been accepted as reasonable one.

On the other hand, Gerl and Zippelius (1999) performed the
 microscopic simulation of the two-dimensional collision of an
 isothermal elastic disk
 with a wall.  Their simulation is mainly based on the mode expansion of 
an elastic disk under the force free boundary condition. 
The distinct characteristic of their model is that they do not introduce 
 any dissipative mechanism of their microscopic equation of motion.
Then, they
 solve Hamilton's equation determined by the elastic field and the
 repulsive potential to represent the collision of two disks.
Their results show that COR decreases with the impact velocity, which
strongly depends on Poisson's ratio. For high velocity of the impact 
they demonstrate the macroscopic deformation left after the
collision is over. 
The relation between 
 the quasi-static theory of impact 
(Kuwabara \& Kono, 1987; Brilliantov et al., 1996; Morgado \& Oppenheim, 1997) 
and their microscopic
 simulation (Gerl \& Zippelius , 1999) is not trivial,
because the energy transfer during the impact is not explicitly included in
the quasi-static theory.
Thus, we have to clarify the relation between two typical approaches.

In this paper, we will perform the microscopic simulation of the impact 
of a two dimensional elastic disk with a wall. We introduce two methods
of simulation: one is based on the lattice model (model A)
and another is a continuum model (model B) which is identical to that 
by Gerl and
Zippelius (1999). Both models do not include any dissipation
explicitly.
Thus, we regard inelastic collisions take place only from the transfer of 
modes of oscillation.
Through our simulation, we will demonstrate that 
 the elastic models do not  recover the results predicted by 
the quasi-static
theories in the low impact velocity (Kuwabara \& Kono 1987;
 Brilliantov et al., 1996; 
Morgado \& Oppenheim, 1997; Schwager et al., 1998; Ram\'irez et al.,
1999).


The organization of this paper is as follows. In the next section, we
will briefly review the outline of 
quasi-static theory (Kuwabara \& Kono, 1987; Brilliantov et al., 1996;
 Morgado et al., 1997). 
In section 3, we will explain model A and model B which is equivalent to 
the model by Gerl and Zippelius (1999) of our simulation.
In section 4, we will show the result of our simulation and discuss the
validity of quasi-static theory. 
In section 5, we will discuss our results.
 In section 6, we will summarize our result.
In Appendix, we summarize the outline of the quasi-static theory by
 Morgado and Oppenheim (1997). Note that
parts of this paper has been published as  separated papers
(Hayakawa \& Kuninaka.,2001a;
 Hayakawa \& Kuninaka, 2001b; Kuninaka \& Hayakawa, 2001). 

\section{Quasi-Static Theory: Review}

In this section, we briefly explain the outline of the quasi-static theory.
One purpose of this section is to summarize the two-dimensional version
of quasi-static theory which may not be mentioned in any articles
explicitly. 

At first, let us summarize the three dimensional result, in which
the equation of the macroscopic deformation is given by
\begin{equation}\label{oppenheim}
\ddot h=-k_h( h^{3/2}+A_h \sqrt{h}\dot h)
\end{equation}
in a collision of two spheres. For the collision of two identical spheres
 the macroscopic deformation $h$ is given by
 $h=2R-|{\bf r}_1-{\bf r}_2|$ with the radius $R$ and the position
of the center of the mass ${\bf r}_i$ of $i$ th
particle. $\dot h$ and $\ddot h$ are respectively $dh/dt$ and $d^2h/dt^2$.
$k_h$ in eq.(\ref{oppenheim}) is written as
\begin{equation}\label{eq3}
 k_h=\frac{\sqrt{2R}Y_0}{3(1-\sigma^2)M_0}
\end{equation}
 where $M_0$, $Y_0$ and $\sigma$ are the mass of the sphere,
(three dimensional) Young's modulus, and Poisson's ratio, respectively. 
In eq.(\ref{oppenheim}) $A_h$ is a constant, which may be a function of
viscous parameters (Brilliantov et al., 1996).
 The first term of the right
hand side in eq.(\ref{oppenheim})
 represents the Hertzian contact force 
(Love, 1927; Landau et al., 1960; Johnson, 1985; Hertz, 1882)
 and the second term is
the dissipation due to the internal motion.  

The simplest derivation of eq.(\ref{oppenheim})
 is that by Brilliantov {\it et
al.} (1986), though we
also check its validity by the alternative methods. Taking into account
the limitation of the length of this paper, we follow the argument by
them. The outline of the derivation by Morgado and Oppenheim (1997) 
is shown in 
 Appendix.

The static stress tensor in a two-dimensional 
linear elastic material can be represented by
\begin{equation}\label{ela}
{\sigma^{(el)}}_{ij}=2\mu (u_{ij}-\delta_{ij}u_{ll}/2)+K\delta_{ij}u_{ll}
\end{equation}
where $\mu$ and $K$ are respectively the shear modulus 
and the bulk modulus, and $u_{ij}$ is given by 
\begin{equation}
u_{ij}=\frac{1}{2}\left(\frac{\partial u_i}{\partial x_j}+
\frac{\partial u_j}{\partial x_i}\right)
\end{equation}
with the displacement field $u_i$. 

 The two dimensional Hertzian contact law (Johnson, 1985; Gerl et al., 1999)
 is given by
the relation between the macroscopic deformation of the center of mass
 $h$ and the elastic force $F_{el}$  as
\begin{equation}
h\simeq-\frac{F_{el}}{\pi Y}\{ \ln \left(\frac{4 \pi Y
        R}{F_{el}\left(1-\sigma^{2}\right)}\right)-1-\sigma\} ,\label{eq:hertz}
\end{equation}
where $Y$ is (two-dimensional) Young's modulus.
Note that $F_{el}$ and $Y$ do not have dimension of the force and 
Young's modulus, because these are 
two dimensional variables
 which are the ones per unit length along the third axis.
Equation (\ref{eq:hertz}) can be derived from the stress tensor
(\ref{ela}) with the standard treatment of linear elastic theory.
Note that $h$ satisfies $h=R-y_0$ with  the position of the
 center of mass $y_0$ (Gerl \& Zippelius, 1999). 

For small dissipation, as in Landau \& Lifshitz (1960),
 the dissipative 
stress tensor due to the viscous
motion among internal motions is given by
\begin{equation}\label{vis}
{\sigma^{(vis)}}_{ij}=2\eta_1 (\dot u_{ij}-\delta_{ij}\dot u_{ll}/2)+\eta_2\delta_{ij}\dot u_{ll},
\end{equation}
where $\dot u_{ij}$ is the time
derivative of $u_{ij}$, $\eta_{i}$ ($i=1,2$) is the viscous
constant.

Brilliantov {\it el al.} (1996)
 assumed that the velocity of deformation field is
governed by the macroscopic deformation, {\it i.e.},
$\dot u_i\simeq \dot h(\partial
u_i/\partial h)$. 
Since in the limit of $v_i\to 0$  we may replace eq.(\ref{eq:hertz})
by $F_{el}\simeq -\pi Y h/\ln(4R/h)$ (Gerl \& Zippelius, 1999). 
Thus, with the aid of the assumption by Brilliantov {\it et al.}
 (1996), (\ref{ela}) and (\ref{vis}), it is easy to derive
the two dimensional version of quasi-static theory  as
\begin{equation}\label{2d-quasi} 
F_{tot}\simeq -\frac{\pi Y  h}{\ln(4R/h)}- A \frac{\pi Y \dot h}{\ln(4R/h)} , 
\end{equation}
where $A$ is not an important constant. 
This result can be derived by various other methods. 
In section 4,
 we will compare the result of our simulation with eq.(\ref{2d-quasi}).

\section{Our Models}

Let us explain the details of our models to simulate collisions between
two identical disks whose radius $R$ by the method of the mirror image. 
In both models, the 
wall exists at $y=0$, and the center of mass of the disk keeps the
position at $x=0$. The disk approaches from the region $y > 0$
 before rebounding from the wall.

\subsection{Model A}
The disk in model A
 consists of some mass points (with the mass $m$) 
on a triangular lattice. All the mass points are connected
 with linear springs with spring constant $\kappa$.
 In the limit of a large number of mass points, this disk
 corresponds to the continuum circular disk with 
  Young's modulus 
  $Y=2 \kappa / \sqrt{3}$ and Poisson's ratio $1/3$ (Hoover, 1991).
 The position of each mass point of model A is governed by the
 following equation:
\begin{equation}\label{modelA}
m \frac{d^2 {\bf r}_p}{dt^2}=-\kappa\sum_{i=1}^6(d_0-|{\bf r}_p-{\bf r}_i|)
\frac{{\bf r}_p-{\bf r}_i}{|{\bf r}_p-{\bf r}_i|}
+{\bf e}_y a_0 V_{0} e^{-a_0 y_{p}}
\end{equation}
 where $d_0$ is the lattice constant, ${\bf r}_{i}$ is the
 position of the nearest neighbor mass points of
 ${\bf r}_{p}$, $m$ is the mass of the mass points,
 $y_{p}$ is the $y$ coordinate of ${\bf r}_{p}$, and
 ${\bf e}_y$ is the unit vector in the $y$ direction.
 Note that the directional projection of the linear spring force in model A
can cause the nonlinear deformation. 
The
 wall potential is  
given by
$V_{0}e^{-a_0 y} $
, where $V_{0}=mc^{2}a_0d_0/2$ with $c=\sqrt{Y/\rho}$ and the density $\rho$. 
We adopt  $a_0=100 /d_0$  for the most of simulations, but we also adopt 
the result of
 $a_0=25/d_0=500/R$ 
to obtain Fig.2, though the result is almost identical to that for 
$a=100/d_0$. 
 The exponential interaction between the disk and the wall is
 introduced to simulate a collision between two identical
 disks. Actually,
in the limit of $a_0\to \infty$, the exponential potential can be
regarded as a  potential of the mirror image. Thus, for later
calculation, we analyze the case for large $a_0d_0$. 
The number of mass points is fixed at 1459 in model A, since the rough
 evaluation of convergence of the results has been checked in this model.

\subsection{Model B}

In this subsection, we introduce model B which is originally proposed by 
 Gerl and Zippelius (1999). Although the details of this model can
 be found in their paper, we present a short description of this
 model to understand the setup of our simulation.

 Gerl and Zippelius (1999)
 analyze Hamilton's equation to simulate collisions of a disk with the
 radius $R$ as;
\begin{equation}\label{H's equation}
\dot P_{n,l}=-\frac{\partial H}{\partial Q_{n,l}} ;\quad
\dot Q_{n,l}=\frac{\partial H}{\partial P_{n,l}}
\end{equation}
under the Hamiltonian
\begin{equation}\label{hamiltonian}
H=\frac{P_0^2}{2M}+\sum_{n,l}^N(\frac{P_{n,l}^2}{2M}+\frac{1}{2}M\omega_{n,l}^2
Q_{n,l}^2)+V_1\int_{-\pi/2}^{\pi/2}d\phi e^{-a_0y(\phi,t)} .
\end{equation}
Here $M$ is the (two-dimensional) mass of an elastic disk, and 
$Q_{n,l}$ is the expansion coefficient of the 2D elastic deformation 
field  in the polar coordinate ${\bf u}=(u_r,u_{\phi})$ 
\begin{equation}
(u_r(r,\phi),u_{\phi}(r,\phi))=\sum_{n,l}
Q_{n,l}(u_r^{n,l}(r)\cos n\phi,u_{\phi}^{n,l}(r)\sin n\phi),
\end{equation}
where $u_r^{n,l}(r)R=A_{n,l}\displaystyle\frac{dJ_n(k_{n,l}r)}{dr}+
n B_{n,l}\displaystyle\frac{J_n (k'_{n,l}r)}{r}$
and 
$u_{\phi}^{n,l}(r)R=-n A_{n,l} \displaystyle\frac{J_n(k'_{n,l}r)}{r}-
B_{n,l}\displaystyle\frac{dJ_{n,l}(k_{n,l}r)}{dr}$ 
with the radius of the disk and the Bessel function of the $n-$th order $J_n(x)$.
Here $k'_{n,l}=k_{n,l}\sqrt{2(1+\sigma)/(1-\sigma^2)}$ and $k_{n,l}$ is 
the solution of
\begin{eqnarray}
& &(1-\sigma^2)(1-n^2)\kappa\kappa'^2J_{n-1}(\kappa)J_{n-1}(\kappa')
+ \kappa^2[\kappa^2-2n(n+1)(1-\sigma)]J_n(\kappa)J_n(\kappa')
\nonumber \\
& & + (1-\sigma)[\kappa^2-(1-\sigma)(1-n^2)n][\kappa J_{n-1}(\kappa)J_n(\kappa')
+\kappa'J_{n-1}(\kappa')J_n(\kappa)]=0
\end{eqnarray}
with Poisson's ratio $\sigma$, $\kappa=k_{n,l} R$ and
$\kappa'=k'_{n,l}R$, which is given by the force free boundary
condition of the disk:
\begin{equation}\label{free}
\sigma_{r\phi}(R,\phi)=0
\end{equation}
Thus, for fixed $n$ there are infinitely many solutions $k_{n,l}$
and $\omega_{n,l}=k_{n,l}\sqrt{Y/\{\rho(1-\sigma^2)\}}$
 numbered by
$l=0,1,\cdots,\infty$.
$A_{n,l}$ and $B_{n,l}$ are
determined by 
\begin{eqnarray}
& & -A_n[\frac{(1-\sigma)}{R}\frac{dJ_n(k_{n,l}R)}{dR}+
(k_{n,l}^2-\frac{(1-\sigma)}{R^2} n^2)J_n(k_{n,l}R)] \nonumber \\
& &+n B_n(1-\sigma)[
\frac{1}{R}\frac{dJ_n(k'_{n,l}R)}{dR}-\frac{J_n(k'_{n,l}R)}{R^2}]=0
\end{eqnarray}
 and
$\int_0^Rdr r \{{u_r^{n,l}}^2+{u_{\phi}^{n,l}}^2\}=R^2$. 
 $P_{n,l}$ is
 the canonical momentum.
$y(\phi,t)$ is the shape of the elastic disk in polar
coordinates;
\begin{equation}
y(\phi,t)=
y_0(t)+
\sum_{n,l}Q_{n,l}(C_{n,l}\cos(n\phi)\cos\phi-S_{n,l}\sin(n\phi)\sin\phi)
\end{equation}
with the position of the center of mass $y_0(t)$ and constants $C_{n,l}$ 
and $S_{n,l}$ determined by the maximal radial and tangential
displacement at the edge of the disk as $C_{n,l}=u_r^{n,l}(R)$ and 
$S_{n,l}=u_{\phi}^{n,l}(R)$.
 $M$ is the mass of the disk, and the momentum of the
center of the mass $P_0=M\dot{y_{0}}$ satisfies 
$ \dot P_0=- (\partial H/\partial y_0)$
, $V_0$ and $a$ are parameters to express
 the strength of the wall potential.

For the simulation of a pair of identical disks, they
have confirmed that the result with finite $a_0$ can be extrapolated to
the result of $a_0\to \infty$ by taking into account finite $a_0$ effect 
in proportion to $1/(a_0R)$. Similarly,
the result with finite number of modes $N$ 
should  be extrapolated 
with the correction in proportion to $1/\sqrt{N}$. Since they have
already checked such the tendencies, 
we only adopt $N=1189$ ($n\le 50$ and $\kappa_n\le 50$)or $N=437$
($n\le 30$ and $\kappa_n\le 30$), 
$V_1=Mc^2a_{0} R/2$ and
 $a_0=500/R$.


\subsection{Parameters in both models}

For the comparison between two
different models, 
we only simulate the case of Poisson's ratio $\sigma=1/3$. 
The numerical integration scheme for model A is the classical
 fourth order Runge-Kutta method with 
$\Delta t=1.6\times 10^{-3} \sqrt{m/\kappa}$. 
Parts of the calculation in model A has been checked by the fourth order 
symplectic integral method with $\Delta t =5.0 \times
10^{-3}\sqrt{m/\kappa}$,
 and no differences in results of two methods can
be found.
For model B,
 we adopt the fourth order 
symplectic integral method with $\Delta t=5.0\times 10^{-3}R/c$.
In both models, we have checked for conservation of
 the total energy. 

 We also investigate the impact with finite 
 temperature. The temperature is introduced as follows:
In model A, we prepare the Maxwellian for the initial velocity
 distribution of mass points, 
where the positions of all mass points are located  at their equilibrium
 positions.
 From the variance of the Maxwellian we can 
 introduce the temperature as a parameter.
 To perform the simulation,
we prepare 10 independent samples obeying Maxwellian with the
 aid of normal random number.
In model B, we prepare samples which satisfies Gibbs states.
Namely, $\sqrt{M}\omega_{n,l}Q_{n,l}/\sqrt{2}$
 and $P_{n,l}/\sqrt{2M}$ obey the normal
 random number with the variance (temperature) $T$.
 In model B, we prepare many samples (120 or 20) 
 to simulate  systems at finite $T$.

The summary of differences between model A and B is as follows:
(i) All of the mass points in model A interact with  the wall
 but,  in model B, 
only the exterior boundary has the influence of the potential as in
 eq.(\ref{hamiltonian}). We have replaced the original model A by a
 model in which only mass points on the boundary can interact with the
 wall, but we cannot find significant differences in the results of our
 simulation in both discrete models. 
(ii) Model A can have nonlinear deformations,
 but model B is based on the theory
 of linear elasticity. 
(iii) Model A can express some plastic deformations, but model B cannot.
This effect will be discussed in section 6.
(iv) Model A has six fold symmetry whereas model B has only rotational
 symmetry. 
(v) The force free boundary condition (\ref{free}) is assumed in model B 
but may not be appropriate for actual situations. Model A does not
include such the condition. 
   
\section{Results}


Now, let us explain the details of the result of our simulation.
In the first subsection, we will introduce the result at $T=0$ and in
the second subsection, we will show the result at finite $T$.

\subsection{Simulation at $T=0$}
 
At first, we carry out the simulation of model A and model B 
with the initial condition at $T=0$ ({\it i.e.} no internal
motion).
Figure 2 is the plot of 
the COR against
the impact velocity for both model A and model B. 
For model A, we have adopted the fourth order Runge-Kutta method.
 To eliminate the effect of six fold symmetry of model 
A, we average 12 data as a function $\theta$ of the initial orientations 
of the disk i.e. $\theta=\pi n/72$ with $n=1,2,3,\cdots,12$
with $a_0=25/d_0=500/R$ for $N=1459$. 
We also investigate the case that only mass points at the boundary can
interact with the wall for small $v_i$ but their results do not have any 
visible difference from the original model A. It is obvious that there
is no plastic deformations for $v_i\le 0.2c$.

For
  model B, we show the results of $437$ modes and $1189$ modes which
  clearly demonstrates the convergence of the result for the number of modes.
 When impact velocity $v_{i}$ is larger than $0.1c$ with $c=\sqrt{Y/\rho}$,
 the value of COR of
 model A is almost identical to that of model B.
 Each line decreases smoothly as impact velocity increases.

At present, we do not know the reason why the significant difference
 between the two models exists at low impact velocity. 
It is difficult to imagine that occurrence of  nonlinear deformations during
the impact of model A causes the difference
because the deformation is smaller when  $v_i$ is smaller.


Second,
 we investigate the  force acting on the center of mass of the disk
caused
 by the interaction with the wall in model B.
In the limit of $v_i\to 0$ we expect that the Hertzian contact 
theory can be used(Landau \& Lifshitz, 1960;
 Johnson, 1985; Gerl \& Zippelius, 1999).
The small amount of transfer from the translational motion
 to the internal motion is  the macroscopic dissipation.
Thus, we  can check whether the quasi-static 
approaches (Kuwabara \& Kono, 1987; Brilliantov et al., 1996;
 Morgado \& Oppenheim, 1997)
 or our elastic simulation can be used in slow impact situations.

 
If $h$ is given, we can calculate the
 elastic force by solving eq.(\ref{eq:hertz}) numerically.
Figure 3 is the comparison with
our simulation in model B (1189 modes) and the Hertzian contact theory 
(\ref{eq:hertz}) which is given by
 the solid lines.
 The result of our simulation at the impact velocity 
$v_i=0.01c$  shows the hysteresis as 
suggested in the simulation at $v_i=0.1c$ (Gerl \& Zippelius, 1999).
This means the compression and rebound are not symmetric.
The hysteresis curve is still self-similar even at
 $v_i=0.04c$ but the loop becomes noisy at $v_i=0.1c$.

For very low impact velocity $v_i=0.001c$, the hysteresis loop
almost disappears and the total force observed in our simulation is almost a
linear function of $h$ which deviates from the one predicted by
both the Hertzian contact theory
and the quasi-static theory (\ref{2d-quasi}).
In particular, the turning point which corresponds to the point of the largest
$F_{tot}$ in Fig. 3(b) is apart
from the Hertzian 
curve (the solid line). This deviation is in clear contrast to the quasi
static theory, because the dissipative force in the theory
in eqs.(\ref{oppenheim}) and (\ref{2d-quasi}) 
must be zero at the turning 
point which $\dot h=0$ should satisfy.
This tendency is invariant even for the simulation of model A, though
the data becomes noisy. 
The linearity of the total repulsion force is not
surprising, because
$e^{-a_0 y(\phi,t)}$ in the potential term in
eq.(\ref{hamiltonian}) can be expanded in a series of 
$Q_{n,l}$ for very slow impact.


The result may suggest that our elastic models do not recover the Hertzian
contact theory in the quasi-static limit. To check the tendency, we
investigate whether any static state can be achieved in our models in the 
compression. Figure 4 is the time evolution
of the center of mass in the simulation of model B, where the strength
of dimensionless external field  is $g=0.01c^2/R$. We observe that 
an undamped harmonic oscillation of the center of mass 
in the simulation after
the first deformation. This oscillation is stable because the energy
of oscillation is not enough to overcome  finite energy gap between
energy levels. Thus, the
center of mass keeps the oscillation as the motion in the ground state.  
We note that Fig.4 is the result
of the simulation at finite temperature in which the mode transfer is
enhanced. Nevertheless, the center of mass keeps the harmonic oscillation.
This
tendency can be observed in model A, too. 
Even when we introduce the randomness in the coupling in model A, 
the oscillation is undamped.
Thus, both of elastic models
cannot reach any equilibrium steady state as is assumed in the Hertzian
contact theory.  
This result indicates that the elastic models are not
appropriate to describe quasi-static situations for $v_i/c\ll 1$. 
Note that the introduction of nonlinear deformation may not be
enough, because as we can see in Fig.3
 (b) the deformation is
very small for slow impact. Thus, it is difficult to imagine the impact
produces nonlinear deformations.  
To reach an equilibrium state, thus, we need to introduce some microscopic
dissipative mechanism.

However, the validity of the contact time $\tau$ in
the impact evaluated  as $\tau \simeq (\pi R/c) \sqrt{\ln(4c/v_{i})}$ by
the quasi-static theory (Gerl \& Zippelius, 1999) has been
confirmed by the results of our simulation of model A (Fig.5). 
Thus, our elastic model can be valid in the impact with the intermediate 
speed.


\subsection{Simulation at finite $T$}

Now, let us show the results of our simulation at finite $T$.
The thermal velocity 
$v_{th}=\sqrt{T/M}$ 
causes
significant differences from those at $T=0$ in both low and large impact 
velocities. In this sense, we have much 
room to study this process at finite $T$ systematically.


For small impact velocity, i.e. if the effect of $v_{th}$ is not negligible,
  the fluctuation of 
COR at finite $T$ becomes large, while the average is almost independent 
of temperature as in Figs. 6 and 7, where the results are obtained from
  the average of 120 independent samples.
 In some trials at high temperature, thus,  COR becomes larger than 1,
 though the average is less than 1. Of course, for such the high
 temperature, it is impossible to control the actual speed of impact.



For large impact velocity, $v_i\gg v_{th}$,
 we do not observe any definite temperature effect in
model B but we find drastic decrease of COR in model A.
It seems that COR can be on a universal curve when the impact velocity
is scaled by the critical velocity 
above which the COR decreases abruptly (Fig.8). 
 The relation between the critical velocity and the initial 
 temperature at the intermediate impact velocities 
is shown in Fig. 9.
 The critical velocity seems to obey a linear function of $T$,
though the data is not on the function for both 
 slow and  fast impacts. 

\section{Alternative Quasi-Static Theory: The Effect of Temperature Gradient}

In this section, let us discuss new aspects of the
quasi-static theory. As in section 2, the conventional quasi-static theories
(Kuwabara \& Kono, 1987; Brilliantov et al., 1996)
 consider the effect of internal friction.  
Similarly, the Langevin approach (Morgado \& Oppenheim, 1997)
 gives the identical result to 
that by conventional one. In both approaches, it is assumed that the
temperature in disks 
is uniform. However, this assumption is not accurate. It is
known that the rise of temperature is proportional to the divergence of
elastic deformation (Landau \& Lifshitz, 1960).
 Thus, the temperature cannot be uniform.

In this section, we
will evaluate the dissipation rate due to the thermal diffusion 
and show that the contribution of this term is dominant in quasi-static
situations. 
The result may not be complete but meaningful to indicate  
the importance of the thermal diffusion.


In a quasi-static collision, the compression is proceeded in an
adiabatic process. The adiabatic condition is written as 
 $S_0(T)+K\alpha u_{ii}=S_0(T_0)$, where $S_0$, $K$, $\alpha$ and $T_0$
 are respectively the entropy (divided by the Boltzmann constant),
 the bulk modulus, the thermal expansion
 rate and the temperature without any deformation (Landau, 1960).
 From the expansion of the entropy around $T_0$ we obtain
\begin{equation}\label{a1}
T-T_0=-\frac{T_0K_{ad}\alpha}{C_p}u_{ii}=
-\frac{T_0\alpha\rho}{C_p}(c_l^2-c_t^2)u_{ii},
\end{equation}
where $K_{ad}$, $C_p$, $c_l$ and $c_t$ are the bulk modulus in the
adiabatic process, the heat capacity at constant pressure, the sound
velocity of the longitudinal mode and the sound velocity of the tangential
mode, respectively (Landau \& Lifshitz, 1960).
To obtain the final expression we use the two-dimensional relations 
$K_{ad}=Y/(2(1-\sigma))$, $c_l=\sqrt{Y/(\rho(1-\sigma^2))}$ and
$c_t=\sqrt{Y/(2\rho(1+\sigma))}$.
There is the relation between the stress tensor and 
the divergence of deformation $u_{ii}$ as
\begin{equation}\label{a2}
u_{ii}=\frac{1-\sigma}{Y}\sigma_{ii}=\frac{1-\sigma}{Y}(\sigma_{xx}+\sigma_{yy})
\end{equation}
in the two-dimensional elastic medium (Landau \& Lifshitz, 1960). 
Substituting (\ref{a2}) into (\ref{a1}) we obtain
\begin{equation}\label{a3}
T-T_0= \frac{T_0\alpha}{2C_p}\sigma_{ii}. 
\end{equation}
Thus, if $\sigma_{ii}$ is a function of the position, the temperature
field is not uniform, which is contrast to the assumption in previous
quasi-static theory.

It is known that the thermal diffusion causes energy dissipation.
The dissipation rate is given by
\begin{equation}
\label{a4}
\dot E=-\frac{\kappa_T}{T_0}\int d^2{\bf r}(\nabla T)^2, 
\end{equation}
where $\kappa_T$ is the thermal conductivity (Landau \& Lifshitz, 1960). 
The integration in
(\ref{a4}) is performed in all region of elastic disks.
Thus, from (\ref{a3}) and (\ref{a4}), the energy dissipation which is
not included in  previous treatments is need to be considered.

Now, let us evaluate the integral (\ref{a4}). For this purpose, we use
the exact solution of two-dimensional Hertzian contact
problem (Hills et al., 1993).
 The explicit stress tensor is given by
\begin{eqnarray}\label{a5}
\sigma_{xx}&=& p_0y\left[2-\frac{s}{\sqrt{1+s^2}}
-\frac{\sqrt{1+s^2}}{s}-\frac{\hat x^2s^3}{(1+s^2)^{3/2}(s^4+\hat y^2)}
\right],\nonumber \\
\sigma_{yy}&=& -p_0\frac{\hat y^3\sqrt{1+s^2}}{s(s^4+\hat y^2)} ,
\end{eqnarray}
where  $\hat x=x/a$ and $\hat y=y/a$ are scaled 
by the contact radius $a$ which is 
given by 
\begin{equation}\label{a6}
a^2=\frac{4F_{el} R(1-\sigma^2)}{Y}
\end{equation}
for the contact of two identical disks. 
Note that $x$ and $y$ are the position in the Cartesian coordinate whose 
origin is the center of the contact area (see Fig. 10).
$p_0$ in (\ref{a5}) is given by
\begin{equation}\label{a7}
p_0=\frac{2F_{el}}{\pi a},
\end{equation}
where  $s$ in (\ref{a5}) is 
\begin{equation}\label{a8}
s^2=\frac{1}{2}\left\{-(1-\hat x^2-\hat y^2)+
\sqrt{(1-\hat x^2-\hat y^2)^2+4\hat y^2}\right\} .
\end{equation}
 From (\ref{a5}) we obtain $\sigma_{ii}$ 
\begin{equation}\label{a5b}
\sigma_{ii}=p_0\left\{2-\frac{s}{\sqrt{1+s^2}}-
\frac{s^3\hat{x}^2}{(1+s^2)^{3/2}(s^4+\hat y^2)}-
\frac{\sqrt{1+s^2}{(s^4+2\hat y^2)}}{s(s^4+\hat y^2)} \right\} ,
\end{equation}
Thus, the dissipation rate (\ref{a4}) can be calculated in principle.


Note that the numerical integration of (\ref{a4}) is not easy, because
(i) the explicit expression is too complicated, (ii) 
the boundary is modified by the compression, and (iii) the parameter
$\hat R\equiv R/a$ is important and is a function of the impact velocity.
Thus, here, we present a rough analytical evaluation of (\ref{a4}) to
capture the characteristics of  this problem. We note that $\sigma_{ii}$ 
becomes simple in some special situations. For example, $\sigma_{ii}$ at 
$x=0$ which is on the axis of symmetry is given by
 (Hills et al., 1993)
\begin{equation}\label{a9}
\sigma^{in}_{ii}\equiv\sigma_{ii}(0,\hat y)=2p_0\left[\hat y-\sqrt{1+\hat y^2} \right].
\end{equation}
On the other hand, the integral representation of $\sigma_{ii}$ 
\begin{equation}\label{a10}
\sigma_{ii}=-\frac{2y p_0}{\pi}\int_{-a}^a 
d\xi \frac{\sqrt{a^2-\xi^2}}{(x-\xi)^2+y^2} ,
\end{equation}
can be approximated by
\begin{equation}\label{a11}
\sigma^{out}_{ii}
\simeq-\frac{2y p_0}{\pi r^2}\int_{-a}^ad\xi\sqrt{a^2-\xi^2}=
-\frac{\hat y p_0}{\hat r^2}
\end{equation}
far from $x=0$.
Here we use $\int_{-a}^ad\xi\sqrt{a^2-\xi^2}=\pi a^2/2$.
 
For the evaluation of (\ref{a4}), 
we distinguish the inner part $|x|<a $ from the outer part $|x|>a$.
 In the inner region, we may
replace $\sigma_{ii}$ by $\sigma^{in}_{ii}$. Thus, 
$(\nabla T)^2$ in the inner
region may be approximated by
\begin{equation}\label{a12}
(\nabla T)^2_{in}\simeq \frac{T_0^2\alpha^2p_0^2}{4C_p^2a^2}
\left(1-\frac{\hat y}{\sqrt{1+\hat y^2}}\right)^2.
\end{equation}
In the outer region we may replace $\sigma_{ii}$ by $\sigma^{out}_{ii}$
because such the approximation can be used in the most of regions in the
quasi-static situation ($\hat R\gg 1$). Thus, $(\nabla T)^2$ in the outer
region may be approximated by
\begin{equation}\label{a13}
(\nabla T)^2_{out}\simeq \frac{T_0^2\alpha^2p_0^2}{4C_p^2a^2\hat r^4}.
\end{equation}
Of course, these assumptions
cannot be used in general. In particular, near the edge $|x|=a$ the
contribution is expected to be large. However, we believe that the
evaluation under the simplified assumption may be useful as the first
step of the analysis.

In the inner region, the integrand is independent of $x$ and the
integrated region may be approximated as a square domain
 $-a\le x\le a $ and $0\le y<2R$.
Thus, $\dot E_{in}$ in (\ref{a4}) 
can be evaluated as follows:
 From 
\begin{equation}\label{a14}
\int_0^{2\hat R}dt(1-\frac{t}{\sqrt{1+t^2}})^2=
2+4\hat R(1-\sqrt{1+\frac{1}{4\hat R^2}})-\tan^{-1}(2\hat R)
\simeq \frac{4-\pi}{2}
\end{equation}
with $\hat R\gg 1$, we obtain
\begin{equation}\label{a15}
\dot E_{in}\simeq -\frac{(4-\pi)\kappa_TT_0\alpha^2F_{el}^2}{\pi^2C_p^2a^2} .
\end{equation}

For the outer region, $\dot E_{out}$ in (\ref{a5}) is
\begin{equation}\label{a16}
\dot E_{out}\simeq -\frac{\kappa_TT_0\alpha^2F_{el}^2}{4\pi{C_p}^2a^2}\int d\hat x\int d\hat y \frac{1}{\hat r^4}.
\end{equation}
The evaluation of the outer region is more complicated, because the
domain can not be approximated by a simple rectangular domain.
 For the evaluation, we
neglect the deformation of shape of the compressed disk. Thus, the shape 
is approximated by a hemi-circle as in Fig. 10. 
It is convenient to introduce the polar coordinate $(\hat r,\theta)$ 
to evaluate
(\ref{a16}).
For a given angle $\theta$ between $x$ axis and $OQ$ in Fig. 10 
 $\hat r$ is between $\hat r_{min}\equiv \overline{OP}/a$ and $
\hat r_{max}=\overline{OQ}/a$. We also introduce $\theta_{min}$ which is the
cutoff angle for $\theta$. Taking into account $\hat R\gg 1$ we can evaluate 
\begin{equation}\label{a17}
\hat r_{max}\simeq 2\hat R \sin \theta : \quad
r_{min}\simeq \frac{1}{\cos\theta} .
\end{equation}
Since these evaluations are approximate, we need to introduce the
lower cutoff of $\theta_{min}$ by the consistency condition 
$\hat r_{max}(\theta_{min})\ge \hat r_{min}(\theta_{min})=1$. Thus,
$\theta_{min}\simeq
a/2R$.
The upper cutoff of $\theta$ is 
$\theta_{max}=\cot^{-1}(1/2\hat R)\simeq \pi/2-1/2\hat R$.
Thus, the integral in (\ref{a16}) can be evaluated as
\begin{eqnarray}\label{a18}
I &\equiv& \int d\hat x\int d\hat y \frac{1}{\hat r^4}\simeq
\int_{\theta_{min}}^{\theta_{max}}d\theta 
\int_{\hat r_{min}(\theta)}^{\hat r_{max}(\theta)}\frac{d\hat r}{\hat r^3} \nonumber \\
&=& -\frac{1}{8\hat R^2}\int_{\theta_{min}}^{\theta_{max}}
\frac{d\theta}{\sin^2\theta}
+\frac{1}{2}\int_{\theta_{min}}^{\theta_{max}}d\theta \cos^2\theta
\simeq \frac{\pi}{8}-\frac{1}{2\hat R}\simeq \frac{\pi}{8},
\end{eqnarray}
where we use
\begin{equation}\label{a19}
\int_{\theta_{min}}^{\theta_{max}}\frac{d\theta}{\sin^2\theta}
=-\frac{1}{2\hat R}+\cot(\frac{1}{2\hat R})\simeq 2\hat R+\frac{2}{3\hat R}
\end{equation}
with $\cot\theta\simeq 1/\theta-\theta/3$ in the limit of $\theta\to 0$, 
and
\begin{equation}\label{a20}
\int_{\theta_{min}}^{\theta_{max}}
{d\theta}\cos^2\theta =\frac{\pi}{4}-\frac{1}{4\hat R}.
\end{equation}
Substituting (\ref{a18}) into (\ref{a16}) we obtain
\begin{equation}\label{a21}
\dot E_{out}\simeq -\frac{\kappa_T T_0\alpha^2F_{el}^2}{32 {C_p}^2a^2}. 
\end{equation}
 From (\ref{a15}) and (\ref{a21}), the total dissipation rate $\dot
E=\dot E_{in}+\dot E_{out}$ is given by
\begin{equation}\label{a23.1}
\dot E=-\gamma_0\frac{\kappa_TT_0Y\alpha^2F_{el}}{{C_p}^2(1-\sigma^2)R}
\end{equation}
where
\begin{equation}\label{a23.2}
\gamma_0=\frac{\pi^2+128-32\pi}{256\pi^2}.
\end{equation}

The result suggests that the dissipation rate by the thermal diffusion
is dominant in quasi-static situations, because the force $F_{el}$
 appears in 
(\ref{a23.1}) exists even in the limit of zero impact velocity, while
internal frictions considered in conventional quasi-static theory
disappears in the limit of zero impact velocity. 

This result, however, predicts a singular behavior of COR. In fact,
 the rough evaluation of the total energy loss $E_{loss}$ 
by heat diffusion during the
impact is proportional to the impact velocity $v_i$, while the
definition of COR by $E_{loss}$ is $E_{loss}=M v_i^2(1-e^2)/2$.  
Thus, COR may be singular for very small impact velocity. We need to
 consider another mechanism to remove such the singularity.
We also need such the analysis for three dimensional situations where 
the stress field becomes simpler than that for two-dimensional cases
 (Hills et al.  1993).

\section{Discussion}


 We investigate what happens in the disk
 above the critical velocity and find the existence of plastic
 deformation of the disk (Fig. 11(a)).
Actually, there are no energy differences between two configurations in
 Fig. 11(b) 
which can occur after the strong compression during the impact
 but cannot be released  after the impact is over. 
It is well known that plastic deformation causes the drop of the 
COR (Johnson, 1985). 

\subsection{Application of the Conventional Theory of Plastic
 Deformation to 2D Impacts
}

Following the description by Johnson (1985), 
let us explain the dimensional analysis of the two-dimensional plastic
deformation. From two-dimensional Hertzian law (\ref{eq:hertz}) we evaluate
     $h \sim a^2/R$ (Johnson, 1985).
 The work for the compression of the
     disk $W$ is $W=(1/2)Mv_{i}^2 \sim
     \int_{0}^{h^{*}}dhF_{el} \sim 
     \int_{0}^{a^{*}}da a^3/R^2$, where $M$ and
     $v_{i}$ are the mass of the disk and the impact
     velocity, respectively. $h^{*}$ and $a^{*}$ are
     respectively the maximal compression and and the
     maximal contact length. Here we neglect the logarithmic 
     correction and unimportant numerical
     factors. Introducing the mean contact pressure during
     dynamical loading $p_{d}$ which satisfies $p_{d} \sim
     F_{el}/a$, $W$ can be evaluated by $W \sim F_{el}h^*\sim 
 p_{d}(a^{*})^{3}/R$. From $W\sim Mv_{i}^{2}$ we can
     express $a^{*} \sim (Mv_{i}^{2}R/p_{d})^{1/3}$.

Let us assume that the impact exceeds the yield pressure for 
the plastic deformation. In such the case, the deformation
during rebound is frozen. Thus, the work in a rebound is $W' 
\sim F^{*}h^{*}$ where $F^{*}$ is the maximal force during
the impact. From $h^{*}\sim F^{*}/Y$ and $F^{*}\sim
p_{d}a^{*}$ we evaluate $W' \sim
(p_{d}a^{*})^{2}/Y$. Substituting the expression of
$a_{0}^{*}$ into the expression for $W$ and $W'$ we obtain
the COR as 

\begin{equation}\label{e-plastic}
e^{2}=\frac{v_{r}^{2}}{v_{i}^{2}}=\frac{W'}{W} \sim 
\frac{p_{d}^{4/3}R^{2/3}}{Y(Mv_{i}^2)^{1/3}} .
\end{equation}
Thus, we expect the law $e \sim v_{i}^{-1/3}$ in the
collision of a plastic deformed disk. The three dimensional
version of evaluation which gives $e \sim v_{i}^{-1/4}$
agrees well with the experiment (Johnson, 1985).

\subsection{Realistic Systems}

The actual plastic deformation is more complicated than what we modeled 
in this paper. For example, in the actual contact area 
a central region of perfect
contact is surrounded by an annulus of imperfect contact. In actual
situations, it is not easy to obtain a pure normal collision, because
the rotation of disks is difficult to be suppressed and the wall is not
perfectly flat. Thus, a little deviation of 
the collision
angle causes the tangential stress in collisions.
 In the existence of
tangential stress, we need to consider the effect of imperfect contact
or partial slip in the outer region to get finite force at the corner of 
contact area.

We also note that the actual materials are not uniform. They contain a
lot of microcracks, and amorphous structure locally.  Such the
imperfection of the materials causes the local achievement of the yield 
of plastic deformation.
Thus, the plastic deformation  also occurs localized in contrast to
the macroscopic deformation in Fig.11.

 Our finding is, however, something new, because (i) the decrease of COR
 is excited by the temperature and (ii) COR decreases more
 rapidly like $e\sim v_i^{-1.2}$ than that for the conventional plastic
 deformation $e\sim v_i^{-1/3}$ in (\ref{e-plastic}).
The mechanism how to occur the plastic deformation is not clear at
 present including the linear law in Fig. 9.

For future refinement of our model to describe plastic deformation, we
need to introduce (i) the initial cracks, (ii) local deformation of
lattices at the initial condition, (iii) the yield of local plastic
deformation or non-Hookian effects of springs, and (iv) porosity
distribution at the initial condition except for the introduction of the 
heat diffusion effects as introduced in section 5. Of course, to compare 
the simulation with experiments, we have to simulate the model in three
dimensional situations.

\section{Conclusion}

We have numerically studied the impact of a two dimensional
elastic disk with
the wall with the aid of model A and model B. 
The result can be summarized as
(i) The coefficient of restitution (COR) decreases with the impact
velocity.
(ii) The result of our simulation is not consistent with the
result of the two-dimensional
quasi-static theory. For large impact velocity, there is hysteresis in
the deformation of the center of mass. For small velocity, there remains the 
inelastic force even at $\dot h=0$.
(iii) The effect of heat diffusion may be important for 
the small impact velocity.
(iv) There are drastic effects of temperature in both small and large
impact velocity.
(v) In particular, for large impact velocity of model A,
 we have found the abrupt
drop of COR above the critical impact velocity by the
plastic deformation. The critical velocity of the plastic
deformation seems to obey a
simple linear function of temperature.

We believe that this preliminary 
report is meaningful to recognize
that physicists have poor understanding of  such the fundamental
process of elementary mechanics.
 We hope that this paper
will invite a lot
of interest in the impact from various view points.
We, at least, have a plan to  study three dimensional impacts to
clarify the relation among
the microscopic simulation, experiments and the quasi-static elastic theory. 

\appendix
\par\noindent
\vspace*{0.3cm}
\par\noindent
{\large{\bf Appendix}}

\vspace*{0.3cm}
\section{Langevin Approach to the Quasi-Static Theory}

In this Appendix let us introduce the derivation of quasi-static theory by
Morgado and Oppenheim (1997).
The characteristics of their derivation is to introduce 'thermal
deformation' explicitly to assist the elastic deformation.
Although they do not mention what the thermal deformation is, it is
the complex combination of inelastic scattering of phonons, electrons
, sound radiation into the air and any other mechanism which cannot be
regarded as the elastic deformation.  
In this Appendix we introduce a simplified version of the derivation of
Langevin equation instead of using the original argument 
(Morgado $\&$ Oppenheim, 1997).
 Note that the argument in this Appendix 
is restricted to one in three dimensional systems.

Thus, the position ${\bf d}_i$ of $i$-th. atom can be written as
${\bf
d}_i={\bf R}_i+{\bf u}_i+\Vec{\rho}_i$,
where ${\bf
R}_i$, ${\bf u}_i$, and $\Vec{\rho}_i$ are respectively
the equilibrium position of the atom $i$, the elastic deformation and the 
'thermal' deformation. Here the terminology of 
 'thermal' deformation means that the deformation cannot be controlled
 or the origin has not been specified from macroscopic point of view.
Such the 'thermal' deformations may be regarded as a not 
important variables which can be treated as random variables. 
Let us  introduce $\Vec{\xi}_i={\bf d}_i-{\bf R}_i$.
In the local rule in this Appendix, suffices $i,j$
represent atoms, and the Greek suffices such as 
$\alpha,\beta$ are components.
In addition, we adopt Einstein's rule for suffices where the duplicated
suffices mean the summation as
$a_{\alpha}b_{\alpha}\equiv\sum_{\alpha}a_{\alpha}b_{\alpha}$.
For small deformation,
$\Vec{\xi}_{i}$ can be written as
\begin{equation}\label{II-3-1}
\Vec{\xi}_{i}\simeq 
{\bf R}_i\cdot
\frac{\partial}{\partial {\bf x}_i}{\bf u}({\bf R}_i)+
\Vec{\rho}_i.
\end{equation}

The potential among atoms can be approximated by a harmonic one near its 
equilibrium position. Thus, 
the harmonic potential for isotropic systems is given by
\begin{equation}\label{II-3-2}
U_0=\frac{\kappa}{2}\sum_{i}\Vec{\xi}_i^2 .
\end{equation}
Note that the original paper(Morgado \& Oppenheim, 1997) 
does not assume isotropic form of $U_0$. Then $U_0$ is replaced by a
second order tensor. Substituting (\ref{II-3-1}) into
(\ref{II-3-2}), 
eq.(\ref{II-3-2}) becomes the combination of three terms:
\begin{equation}\label{II-3-3}
U_0=U_{el}+U_{\phi}+U_{H} 
\end{equation}
The first term of the right hand side of (\ref{II-3-3}) 
is the elastic energy for the deformation which can be written as
\begin{equation}\label{elastic}
 U_{el}=\sum_{i}\{\frac{\lambda}{2}
u_{\alpha\alpha}({\bf R}_i)u_{\beta\beta}({\bf R}_i)
+\mu u_{\alpha\beta}({\bf R}_i)u_{\alpha\beta}({\bf R}_i)\} ,
\end{equation}
where $\kappa R_{\alpha}R_{\beta}=\mu\delta_{\alpha\beta}+
\frac{\lambda}{2}\delta_{\alpha\gamma}\delta_{\gamma\beta}$
Note that in the original paper (Morgado \& Oppenheim, 1997),
 the coefficient of 
$u_{\alpha\beta}u_{\gamma\delta}$ becomes a fourth-order tensor.
 $U_{\phi}$ also includes a constant which is represented by a
second-order tensor.

Here, $\lambda$ and  $\mu$ are Lam\'e's elastic coefficients.
The second term of right hand side of
(\ref{II-3-3}) is given by
\[
U_{\phi}=\kappa\sum_{i,j}R_{i\alpha}u_{\alpha\nu}\rho_{j\nu}, 
\]
which expresses the coupling between the elastic deformation and 
the thermal deformation.
The third term of
(\ref{II-3-3}),
$U_{H}=\frac{\kappa}{2}\sum_{i,j}\rho_{i\alpha}\rho_{j\alpha},
$ is the energy of the thermal deformation. The contribution of this
term is in general smaller than other terms.

The collision of two elastic bodies consists of materials 1 and 2.
The energy is the simple summation of the contribution of two materials.
Let the center of mass of i-th particle ${\bf r}^{(i)}$
($i=1,2$). The interaction during collision appears if 
$r_{12}=|{\bf r}^{(1)}-{\bf r}^{(2)}|<2R$.

Morgado and Oppenheim (1997) assume that the slow mode, the
motion of the center of mass, can be written by the solution of the
Langevin equation.
In the Langevin equation, the elastic force is regarded as a systematic
force, while the force
$-\nabla U_{\phi}$ plays a role of the fluctuating force.
As in the general framework of the fluctuation-dissipation theorem,
 the friction coefficient 
$\zeta(r_{12})$ in the Langevin equation 
is determined by the time correlation function of the
fluctuating force as
\begin{equation}\label{II-3-4}
\zeta(r_{12})=\int_0^{\infty}d\tau<\nabla U_{\phi}\nabla U_{\phi}(\tau)>.
\end{equation}
Here $\nabla U_{\phi}(\tau)$ is retarded one of 
$\nabla U_{\phi}$ by time $\tau$. Thus, we can write
\begin{equation}\label{II-3-5}
\zeta(r_{12})=\kappa^2\sum_{k=1}^2\sum_{i,j}R^{(k)}_{\alpha}R^{(k)}_{\beta}
(\nabla_{12}u^{(k)}_{\beta\gamma})(\nabla_{12}u^{(k)}_{\alpha\delta})
\int_0^{\infty}d\tau
<\rho^{(k)}_{i\delta}\rho^{(k)}_{j\gamma}(\tau)>,
\end{equation}
where $\nabla_{12}=\nabla_{{\bf r}_{12}}$ and the upper suffix 
$(k)$ represents the particles 1,2. Here, $\Vec{\rho}_i$ can be regarded
as the thermal fluctuation as
\begin{equation}\label{II-3-6}
<\rho^{(k)}_{i\alpha}\rho^{(k)}_{j\beta}(\tau)>=
\frac{\tau_v}{\kappa} T\delta_{ij}\delta_{\alpha\beta}\delta(\tau) .
\end{equation}
Thus, we obtain 
\begin{equation}\label{II-3-7}
\zeta(r_{12})= T\frac{\tau_v}{\kappa}\sum_{k=1}^2
R^{(k)}_{\alpha}R^{(k)}_{\beta}
(\nabla_{12}u^{(k)}_{\beta\gamma})(\nabla_{12}u^{(k)}_{\alpha\gamma}). 
\end{equation}

Let us  recall that Kramer's equation for many-body systems 
\begin{eqnarray}\label{II-3-8}
\frac{\partial P(X_t,t)}{\partial t}&=&
\left[\left(-\sum_{i=1}^2 \frac{{\bf p}_i}{M}\cdot \nabla_{{\bf r}_i}
+\sum_{i=1}^2\nabla_{{\bf r}_i}U_{el}\nabla_{{\bf p}_i}\right)\right]
P(X_t,t) \nonumber \\
&+ & 
\left[\frac{1}{2}\sum_{j(\ne k)=1}^2\zeta_{12}\hat{\bf r}_{jk}\hat{\bf r}_{jk}
:\nabla_{{\bf p}_{kj}} 
(
\beta\frac{{\bf p}_{kj}}{M}+\nabla_{{\bf p}_{kj}})
\right] P(X_t,t)
\end{eqnarray}
is equivalent to the Langevin equation, where 
$\beta=1/ T$, $X_t=\{{\bf p}_i(t),{\bf r}_{i}(t)\}$ 
($i=1,2$), $\zeta_{12}=\zeta(r_{12})$,
${\bf p}_{kj}={\bf p}_j-{\bf p}_k$,
$\nabla_{{\bf p}_{kj}}=\nabla_{{\bf p}_j}-\nabla_{{\bf p}_k}$.
 $\hat{\bf r}_{jk}$ is the unit vector from the center of $j$ th. particle
to the center of $k$-th. particle. Here we introduce the average
 $<B>_t$ as $<B>_t\equiv \int dX_t B P(X_t,t)$ for any variable $B$.
Note that we have the relation
\begin{equation}
\frac{d}{dt}<B>_t=\int dX_t B \frac{\partial P}{\partial t}.
\end{equation}
Introducing the relative coordinate
${\bf r}_{12}={\bf r}_2-{\bf r}_1$
approximating that the potential $U_0$ is approximated by
$U_{el}$ we obtain
\begin{eqnarray}\label{II-3-9}
<\dot{\bf p}_{12}>_t&=&-\int dX_t P(X_t,t)
(\nabla_{{\bf r}_1}U_{el}-\nabla_{{\bf r}_2}U_{el})
-\frac{\beta}{M} \int dX_t \zeta_{12}\hat{\bf r}_1\hat{\bf r}_2\cdot 
{{\bf p}_{12}}P(X_t,t) \nonumber \\
&=& <{\bf F}_{el}>_t-\frac{\beta}{M}
<\zeta_{12}\hat{\bf r}_{12}\hat{\bf r}_{12}\cdot 
{{\bf p}_{12}}>_t .
\end{eqnarray}
Note the contribution of the linear momentum  disappears from the
integral by parts.
The elastic force $<{\bf F}_{el}>=-\hat{\bf r}_{12}U'_{el}(h)$ is
nothing but the Hertzian contact force, and
$U_{el}=\frac{k_h}{2}h^{5/2}$.
Thus, the time evolution of the macroscopic deformation $h$
($h$ satisfies $h=2R-r_{12}$.) 
is given by
\begin{equation}\label{II-3-10}
\frac{M}{2}\ddot h=-\frac{5}{4}k_h h^{3/2}-\frac{\beta \zeta(h)}{M}\dot h .
\end{equation}
Therefore, to determine the dissipation is reduced to determination of
the friction constant $\zeta(h)$.

Unfortunately, it is impossible to obtain the exact form of
$\zeta(h)$, because $\zeta$ is determined from the complicated relations 
between the elastic deformation and the thermal deformation.
However, from the consideration of power counting of $h$ it is not
difficult to deduce how
$\zeta(h)$ depends on $h$.
In fact, it is easy to show the scaling 
$u_z({\bf x})\to u_z(\sqrt{\alpha}{\bf x})=\alpha u_z({\bf x})$ 
in the Hertzian contact theory.  Similarly we have
 $u_{\alpha\beta}({\bf x})\to 
u_{\alpha\beta}(\sqrt{\alpha}{\bf x})=\sqrt{\alpha}u_{\alpha\beta}({\bf x})$
and $\partial u_{\alpha\beta}({\bf x})/\partial
h\to \partial u_{\alpha\beta}(\sqrt{\alpha}{\bf x})/\alpha\partial h
=1/\sqrt{\alpha}(\partial u_{\alpha\beta}({\bf x})/\partial h)$.
 From the comparison between
the elastic energy (\ref{elastic}) and $\zeta(h)$ in (\ref{II-3-7}), 
it is easy to understand that the key point is how $u_{\alpha\beta}$ and
 $\partial u_{\alpha\beta}/\partial h$ are scaled by $\alpha$.
 From the discussion here the elastic energy is scaled as
$\alpha^{5/2}$, and thus
$\zeta(h)$ is scaled as $\alpha^{1/2}$.
Thus, we finally obtain
\begin{equation}\label{II-3-11}
\zeta(h)=\frac{5}{2}k' h^{1/2}
\end{equation}
and
\begin{equation}\label{II-3-12}
\frac{M}{2}\ddot h=-\frac{5}{4}k_h h^{3/2}-\frac{5}{2}k''\dot h, 
\end{equation}
where $k''=\beta k'/M$ and $k'$ cannot be determined from this argument.
This result agrees with the result by the viscous 
stress tensor (Kuwabara \& Kono , 1987; Brilliantov et al., 1996).

Note that the derivation is quite different from the previous one
assumed the existence of viscous tensor. Both of derivation assumed that 
the temperature field is uniform, but this assumption is not correct in
general. As discussed in the text, the rise of temperature is directly
related to the compression. Since the compression is not uniform, the
rise of temperature is not uniform.

Two-dimensional quasi-static theory in eq. (\ref{2d-quasi}) 
can be derived by a parallel argument introduced in this Appendix.

\vskip 0.4cm
\noindent{\bf Acknowledgment}

We appreciate S. Sasa, S. Takesue, Y.Oono and H. Tasaki
 for their valuable comments. One of 
the authors(HK) thanks S. Wada, K. Ichiki, A. Awazu, and
M. Isobe for stimulative discussions.

\vspace{5mm}
\noindent {\bf Notation\footnote{
In this paper, we analyze two dimensional systems. As a result many of
quantities have different dimensions from those for 
usual three dimensional ones. We also note that we do not introduce
Boltzmann's constant in the calculation. Thus, $T$ has the same
dimension with $E$, and the entropy becomes dimensionless.
}}

\begin{flushleft}
 \begin{longtable}{@{}ll@{}}

  $a_{0}$ & coefficient of a wall potential, 1/m\\
	
  $a $ & contact radius, m\\

  $A, A_h$ & constants, s \\

  $c$ & one dimensional sound velocity, m/s\\

  $c_{l} $ & sound velocity of the longitudinal mode, m/s\\

  $C_{p} $ & heat capacity at constant pressure, dimensionless\\ 

  $c_{t} $ & sound velocity of the tangential mode, m/s\\

  $d_{0} $ & lattice constant, m\\

  $e $ & coefficient of restitution, dimensionless\\

  ${\bf e}_y$ & the unit vector of y direction. \\

  $E $ & energy (2D), J/m \\

  $E_{loss}$ & energy loss during an impact (2D), J/m \\

  $\dot E $  & the dissipation rate (2D), J/m$\cdot$ s \\

  $\dot E_{in} $  & the dissipation rate in the internal region (2D), 
     J/m$\cdot$ s \\

  $\dot E_{out}$  & the dissipation rate in the outer region (2D), 
     J/m$\cdot$ s \\

  $F_{el} $ & 2D elastic force, N/m\\

  $F_{tot} $ & 2D total force, N/m\\

  $g $ & an external field, dimensionless\\ 

  $h $ & macroscopic deformation, m\\

  $H $ & 2D Hamiltonian, J/m \\

  $J_{n}(x) $ & n-th order Bessel function\\

  $k_h $ & the constant in eq.(\ref{eq3}) \\

  $K $ & 2D bulk modulus, N/m\\

  $K_{ad} $ & 2D bulk modulus in the adiabatic process, N/m \\

  $m $ & mass of a mass point (2D), kg/m\\

  $M $ & mass of an elastic disk (2D), kg/m \\

  $M_0$ & mass of a sphere, kg \\

  $P_{n,l} $ & canonical momentum (2D), kg/s \\

  $p_{0} $ & the constant in eq.(\ref{a7})\\

  $Q_{n,l} $ & canonical coordinate, m\\

  $R $ & radius of the disk, m\\

  $\hat R$ & dimensionless radius (R/a) \\

  ${\bf r}_{i} $ & position of mass point\\

  $\hat{r}_{max} $ & maximum distance from origin, dimensionless\\

  $\hat{r}_{min} $ & minimum distance from origin, dimensionless\\

  $\hat{r} $ & distance from origin, dimensionless\\

  $s $ & defined in eq.(22)\\

  $S_{0} $ & dimensionless entropy, \\

  $t $ & time, s\\

  $T $ & temperature (2D), J/m\\

  $T_{0} $ & temperature without deformation, J\\

  $u_{i} $ & displacement field, m\\

  $u_{ij} $ & strain tensor, dimensionless \\

  $V_{0} $ & coefficient of wall potential, J/m\\

  $v_{i} $ & impact velocity, m/s\\

  $v_{r} $ & rebound velocity, m/s\\

  $v_{th} $ & thermal velocity, m/s\\

  $\hat x$ & dimensionless $x$ coordinate of the position ($x/a$) \\

  $Y $ & Young's modulus (2D), N/m$$\\

  $Y_0$ & Young's modulus (3D), N/m$^2$ \\

  $y_{0} $ & position of center of mass, m\\

  $y_{p} $ & $y$ coordinate of $r_{p}$, m\\

  $y(\phi,t) $ & the shape of an elasic disk, m \\

 $\hat y$ & dimensionless $y$ coordinate of the position ($y/a$) \\

 \end{longtable}
  \end{flushleft}

\noindent {\bf Greek letters}	

\vspace{5mm}
\begin{flushleft}
 \begin{longtable}{@{}ll@{}}
  $\alpha $	& thermal expansion rate (2D), m/N\\

  $\beta $  & $1/T$\\

  $\eta_{i} $ & viscous constant (2D), N$\cdot$ s/m\\

  $\gamma_0 $ & the dimensionless constant in eq.(\ref{a23.2})\\

  $\kappa $ & spring constant, N/m\\

  $\kappa_T$ & thermal condutivity (2D), 1/s \\

  $\mu $ & shear modulus (2D), N/m\\

  $\rho $ & density (2D), kg/m$^{2}$\\

  $\sigma $	& Poisson's ratio, dimensionless\\

  $\sigma^{(el)}_{ij} $ & 2-dimensional static stress tensor\\

  $\sigma_{ii}^{in} $ & stress tensor in the inner region\\

  $\sigma_{ii}^{out} $ & stress tensor in the outer region\\

  $\Delta t $ & time step for simulation, s \\

  $\tau $ & contact time, $s$

  \end{longtable}
\end{flushleft}

\noindent {\bf References}
\vspace{5mm}

\noindent Aspelmeier,T., Giese,G., \&	
 Zippelius,A. (1998).Cooling dynamics of a dilute
\hspace*{5mm}gas of inelastic rods: A many particle simulation. {\it
Phys. Rev. E}, 57,  857-865

\noindent Basile,A.  G., \&  Dumont,R.  S. (2000). Coefficient of
restitution for one-dimensi-
\hspace*{5mm}onal harmonic solids. {\it Phys. Rev.  E}, 61, 
2015-2023. 

\noindent Bridges,F. G.,  Hatzes,A.,  \&  Lin,D.N.C.  (1984).
 Structure, Stability and evo-
\hspace*{5mm}lution of Saturn's rings. {\it  Nature}, 309,
 333-335.        

\noindent Brilliantov,N., Spahn,F., Hertzsch,J.-M.  \&  P\"oschel,T.(1996).
  Model for colli-
\hspace*{5mm}sions in granular gases. {\it Phys. Rev.  E},
 53,  5382-5392.       

\noindent Cundall,P.  A., \&  Strack,O.  D. L. (1979).
 A discrete numerical model for gran-
\hspace*{5mm}ular assemblies.
 {\it  G\'eotechinique}, 29,  47-65.       

\noindent de  Gennes,P.  G. (1999).
 Granular matter: a tentative view. {\it  ibid},
 S367-S382
\hspace*{5mm}and references therein.           

\noindent Gerl,F.,  \&  Zippelius,A.  (1999).
 Coefficient of restitution for elastic disks. {\it Phys.
\hspace*{5mm}Rev.  E},
 59,  2361-2372.        

\noindent Giese,G.,  \&  Zippelius,A.  (1996).
 Collision properties of one-dimensional gran-
\hspace*{5mm}ular particles with
 internal degrees of freedom. {\it Phys. Rev.  E},
 54, 4828-4837   

\noindent Goldsmith,W. (1960).
 {\it Impact : The Theory and Physical Behavior of Colliding
\hspace*{5mm}Solids}
 London:Edward Arnold Publ.        

\noindent Hayakawa,H.,  \&  Kuninaka,H. (2001a).
{\it Simulation of the Impact of Two-dimens-
\hspace*{5mm}ional Elastic Disks} ,
in the Prceedings of the Nineth Nisshin Engineering
\hspace*{5mm}Particle Technology 
International Symposium on 'Solids Flow Mechanism
\hspace*{5mm}and Their
 Applications', 82-95.

\noindent Hayakawa, H., \& Kuninaka, H. (2001b). 
 {\it Coefficient of restitution of elastic
\hspace*{5mm}disk}, the Proceedings of Powders and
 Grains 2001 (edited by Y. Kishino),
\hspace*{5mm}561-564, Rotterdam:A. A. Balkema Publ.       

\noindent Hertz,H. (1882a).
 \"Uber die Ber\"uhrung fester elastische K\"orper.
 {\it J. Reine Angew. 
\hspace*{5mm}Math.}, 92,  156-171.

\noindent Hills,D.   A.,   Nowell,D.,  \&  Sackfield,A. (1993). 
 {\it Mechanics of Elastic Contacts}
\hspace*{5mm}Oxford:Butterworth-Heinemann. 
  
\noindent Hoover,W.  G. (1991).
 {\it Computational Statistical Mechanics}. Amsterdam:Elsevier
\hspace*{5mm}Science Publishers B. V.             

\noindent Johnson,K.  L. (1985).
 {\it Contact Mechanics}. Cambridge: Cambridge University
\hspace*{5mm}Press. 

\noindent Kadanoff,L. P. (1999).
 Built upon sand: Theoretical ideas inspired by granular
\hspace*{5mm}flows.
 {\it Rev. Mod.  Phys.}, 71,  435-444.     

\noindent Kuninaka,H.,  \&  Hayakawa,H. (2001).   
The impact of two-dimensional
 elas-
\hspace*{5mm}tic disk, 
{\it J. Phys. Soc. Jpn.},
 70, 2220-2221 .

\noindent Kuwabara,G.  \&  Kono,K.  (1987).
 Restitution Coefficient in a Collision between
\hspace*{5mm}Two Spheres.
 {\it Jpn. J. Appl. Phys.}, 26,  1230-1233.        

\noindent Labous,L.,   Rosato,A.D.,  \&  Dave,R. N. (1997).
 Measurements of collisional 
\hspace*{5mm}properties of spheres using
 high-speed video analysis.
 {\it Phys. Rev.  E}, 56, 
\hspace*{5mm}5717-5725.    

\noindent Landau,L.  D.,\&   Lifshitz,E. M. (1960). 
 {\it Theory of Elasticity}(2nd English ed.).
\hspace*{5mm}New York:Pergamon.             

\noindent Love,A. E. H. (1927).
 {\it A Treatise on the Mathematical Theory of Elasticity}.
\hspace*{5mm}Cambridge:Cambridge Univ. Press.           

\noindent Mindlin,R.  D. (1949).
 Compliance of Elastic Bodies in Contact.
 {\it J. Appl. Mech.
\hspace*{5mm} Trans. ASME}, 16,  259 See also ref.7.       

\noindent Morgado,W.  A., \&  Oppenheim,I.  (1997).
 Energy dissipation for quasielastic
\hspace*{5mm}granular particle collisions.
 {\it Phys. Rev.  E}, 55,  1940-1945. 

\noindent Newton,I.  (1962).
 {\it Philoshophiae naturalis Principia mathematica}.
 London:W.
\hspace*{5mm}Dawason and Sons. The original one has been published in 1687

\noindent Ram\'irez,R.,   P\  oschel,T.,   Brilliantov,N.  \&
 Schwager,T. (1999). Coefficient of
\hspace*{5mm}restitution of colliding
 viscoelastic spheres. {\it Phys. Rev.  E}, 60,  4465-4472 .       

\noindent Schwager,T.,  \& T.  P\"oschel,T.  (1998).
 Coefficient of normal restitution of
\hspace*{5mm}viscous particles
 and cooling rate of granular gases. {\it Phys. Rev.  E},
 57,  650-
\hspace*{5mm}654.   

\noindent Sondergaard,R.,   Chaney,K.,  \&  Brennen,C.  E. (1990).
 Measurements of Solid
\hspace*{5mm}Spheres Bouncing off Flat Plates.
 {\it Transaction of the  ASME,  Journal of
\hspace*{5mm}Applied Mechanics},
 57, 694-699.   

\noindent Supulver,K.   D.,   Bridges,F.   G.,  \&  Lin,D.  N. C. (1995).
 The Coefficient of
\hspace*{5mm}Restitution of Ice Particles in Glancing Collisions:
 Experimental Results
\hspace*{5mm}for Unfrosted Surfaces. {\it  ICARUS},
 113,  188-199.

\newpage

\noindent Figure 1: A schematic figure of a disk used in model A.

\noindent Figure 2: Coefficient of restitution for normal collision of the 
Model A and Model B as a function of impact velocity, where
 $c=\sqrt{Y/\rho}$ with Young's modulus $Y$ and the density $\rho$.
 437 and 1189 modes are chosen for model B. The error bar in model A
 represents the standard deviation of the data as a function of the
 initial orientation $\theta$.

\noindent Figure 3: The comparison of the Hertzian force in
   eq.(\ref{eq:hertz}) with our simulation at $v_{i}=0.01c$
  (a) and  $v_i=0.001c $(b) at $T=0$ in model B. $F_{tot}$ is the total
 force originated from the interaction with a wall.

\noindent Figure 4: The time evolution of the center of mass of the elastic disk
 under the
 compression by $g=0.01c^2/R$ (model B with N=437).
 Here the dimensionless time is measured
 by $R/c$ and the position of C.M. (center of mass) is measured by the
 diameter of the disk (2R). Simulation is performed at the finite
 temperature $T=10^{-8}Mc^2$ and is averaged over 20 independent samples 
 which start from initial condition satisfying the Gibbs distribution.

\noindent Figure 5: The plot of  contact time versus the impact velocity
 (model A). $R$
represents the radius of the disk, in which $R=40$,$60$,$70$ and $80$
 correspond to the
number of mass points 5815, 13057, 17761 and 23233, respectively.
 The dash-dotted line is fitting curve based on
the quasi-static theory. Here the line represents the fit of $\tau c/R$
 as 3.21758$\sqrt{\ln(4c/v_i)}\sim \pi\sqrt{\ln(4c/v_i)}$.

\noindent Figure 6: The average shift of COR at finite temperature $T=10^{-8}Mc^2$
 as a function of the impact velocity in model B with $N=437$.
 The dotted line indicates one at $e(T)=e(0)$.

\noindent Figure 7: The standard deviation of COR
 $\sigma=\sqrt{<(e-<e>)^2>}$
 at
 $T=10^{-8}Mc^2$ as a function of the impact velocity $v_i$
via model B with $N=437$.

\noindent Figure 8: The relation between the coefficient of restitution and
   the impact velocity rescaled by the critical velocity for
   each temperature. Curves are plotted in the log-log
   scale. The temperature is scaled by $T_{0}=mc^2$ with 
the mass of the mass points $m $.
 Note that the error bars are plotted only in the case 
$T/T_{0}=0.03$ but are the same order even at other $T$ (model A).

\noindent Figure 9: The plot of the initial temperature and the
   critical velocity causing the plastic
   deformation. $v_{cr}/c=a(T/T_{0})+b$ is the fitting curve
   line from the data between $T/T_{0}=0.02$ and $0.05$ (model A).

\noindent Figure 10: The configuration of a compressed disk. The origin is $O$. The angle
 between $OP$ and $x$ axis is $\theta$. The inner region is inside two 
 vertical dashed line, while the outer region is outside the central
 region. The length of OS is equal to $a$. The radius of (undeformed) 
disk is $R$.

\noindent Figure 11: (a) Plastic deformation of model A with $v_{i}=0.22c$
   at $T=0.03mc^2$. The solid circle represents the initial
   circle. The points in a circle are positions of the mass points 
   after the collision. The deformation is asymmetric because of the
 velocity distribution at the initial stage.
(b) All the mass points of the disk initially consist of a triangular
 lattice. When the deformation occurs, it is possible that the
 configuration of mass points (points in figure) 
locally change like this figure. 
Note that these two configurations are energetically equivalent.

\newpage


\begin{figure}[htbp]
 \epsfxsize=12cm
 \centerline{\epsfbox{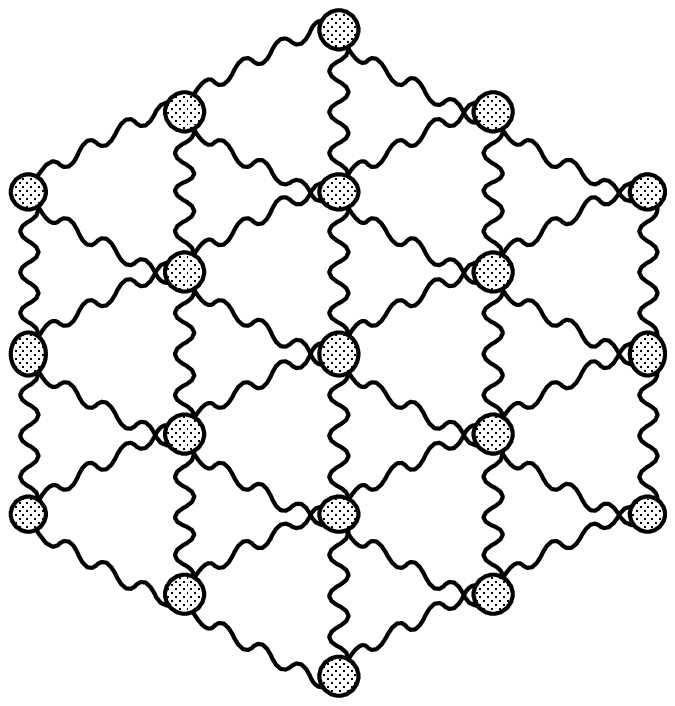}}
 \centerline{\figurename{ 1. authors: Hisao Hayakawa \& Hiroto Kuninaka}}
 \label{network}
\end{figure}

\newpage

\begin{figure}[htbp]
 \epsfxsize=15cm
 \centerline{\epsfbox{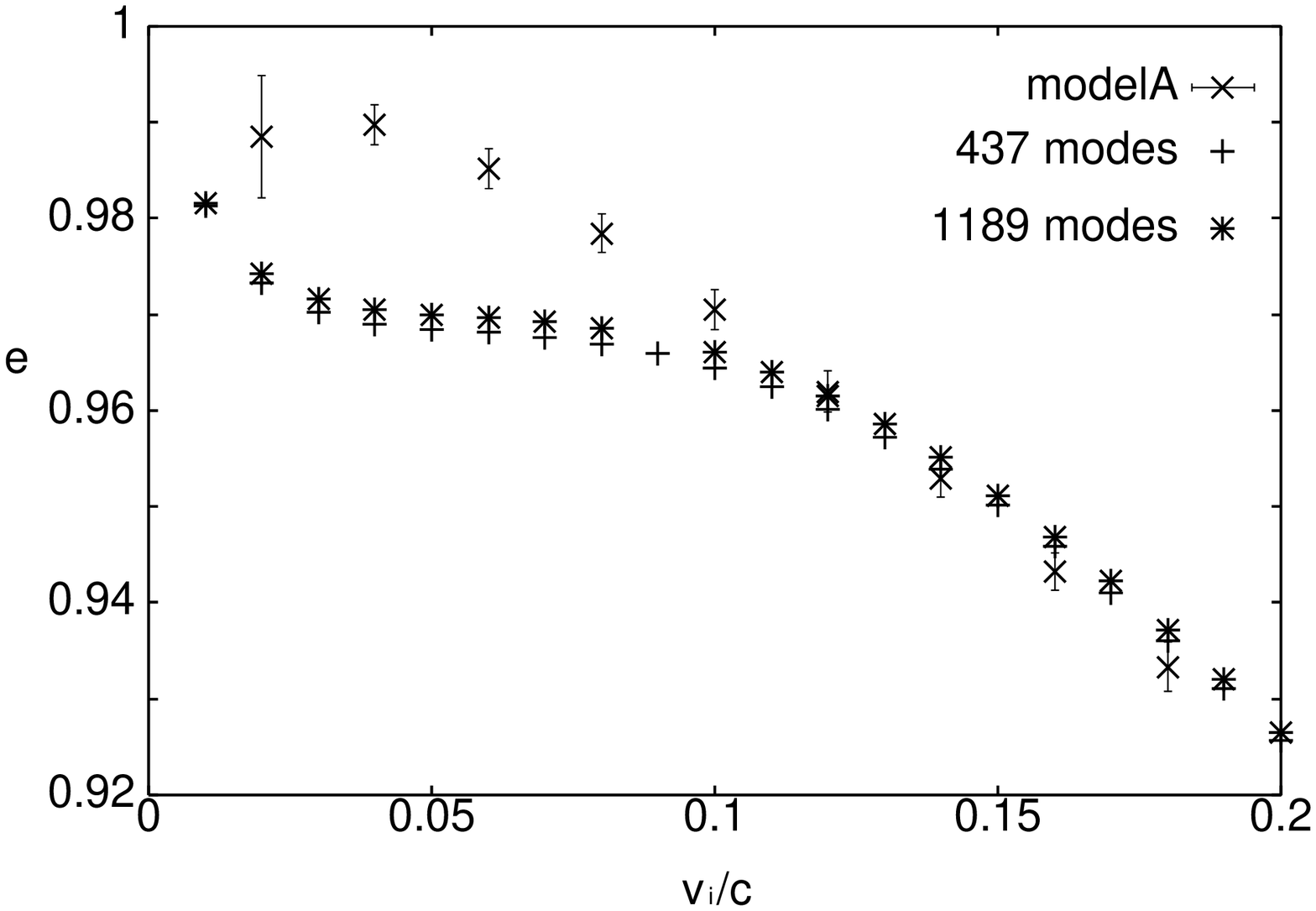}} 
 \centerline{\figurename{ 2. authors: Hisao Hayakawa \& Hiroto Kuninaka}}
 \label{rest}
\end{figure}

\newpage

\begin{figure}[htbp]
 \epsfxsize=12cm
 \centerline{\epsfbox{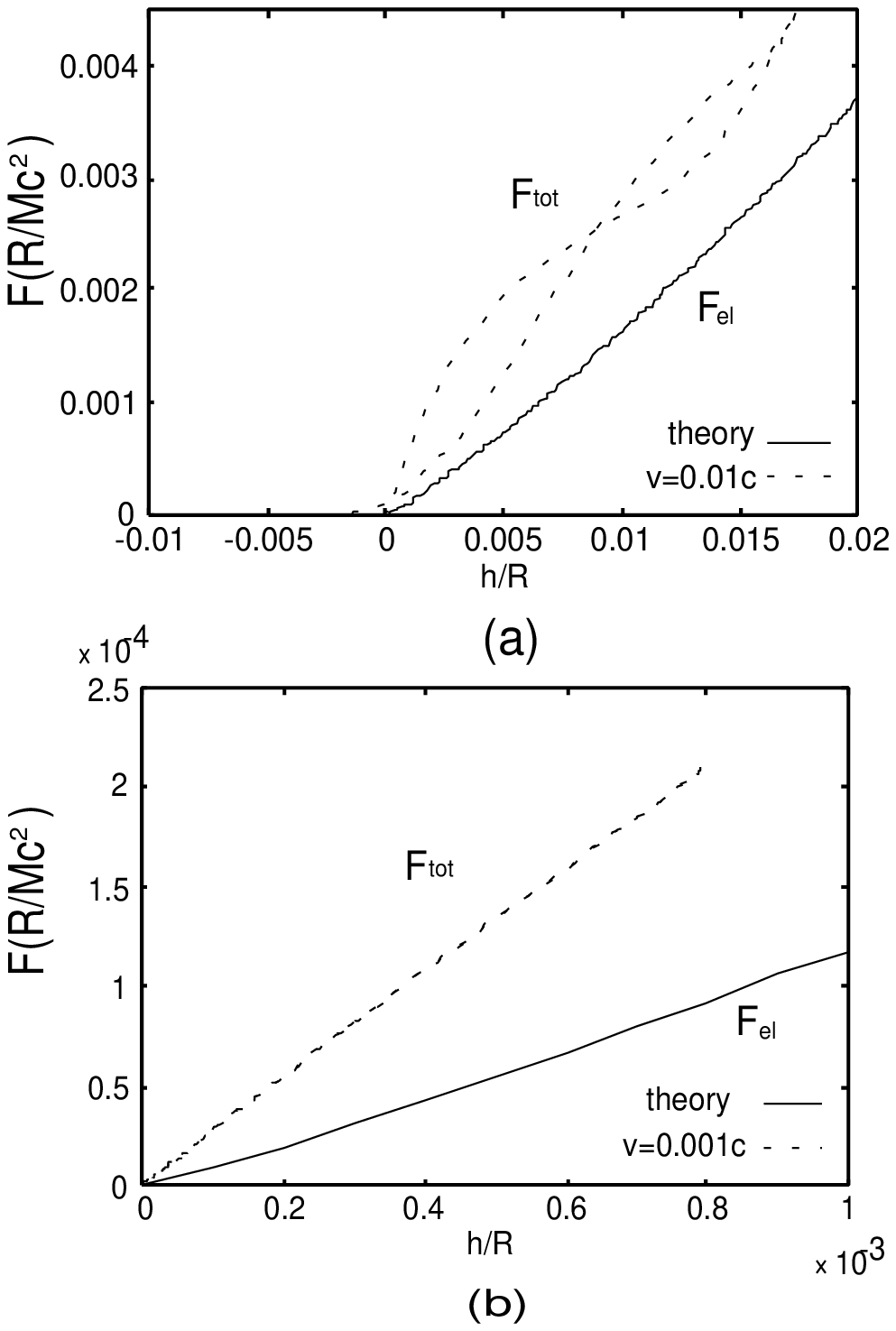}}
 \centerline{\figurename{ 3. authors: Hisao Hayakawa \& Hiroto Kuninaka}}
 \label{fel2}
\end{figure}

\newpage

\begin{figure}[htbp]
 \epsfxsize=20cm
\centerline{\epsfbox{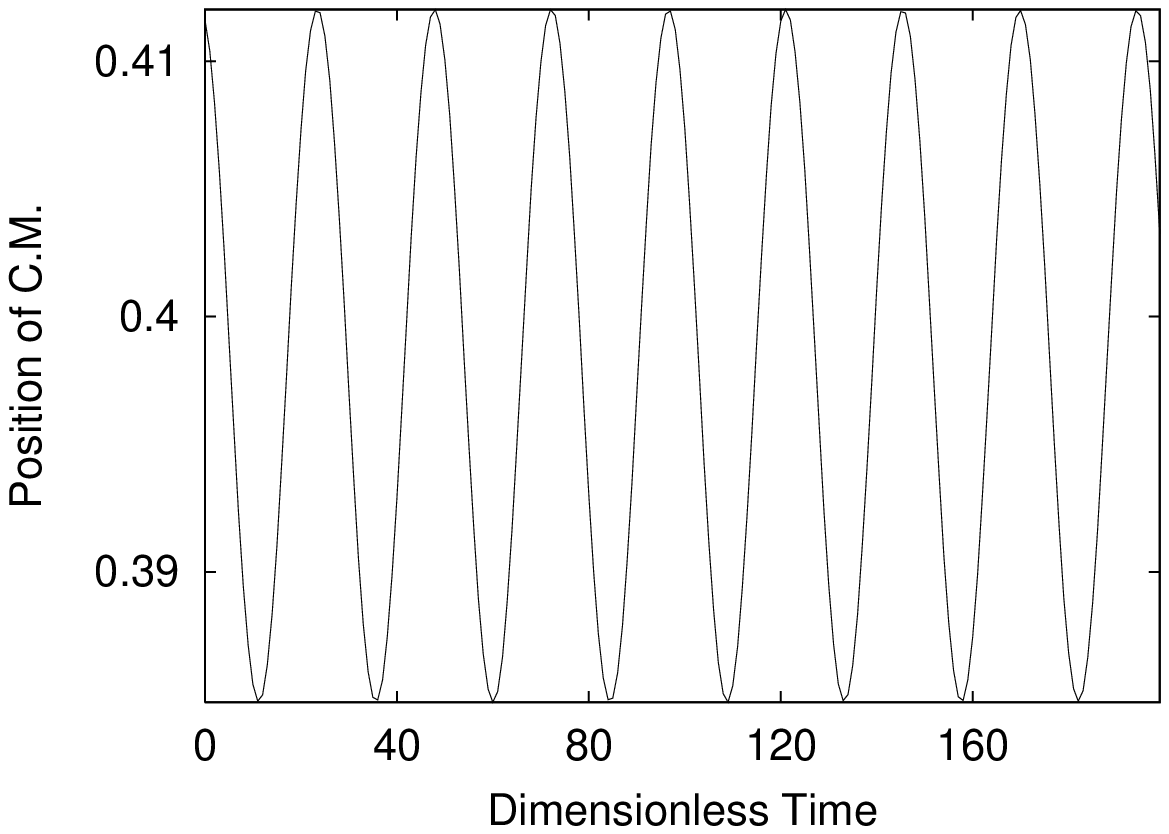}}
 \centerline{\figurename{ 4. authors: Hisao Hayakawa \& Hiroto Kuninaka}}
\label{osci}
\end{figure}

\newpage

\begin{figure}[htbp]
 \epsfxsize=20cm
\centerline{\epsfbox{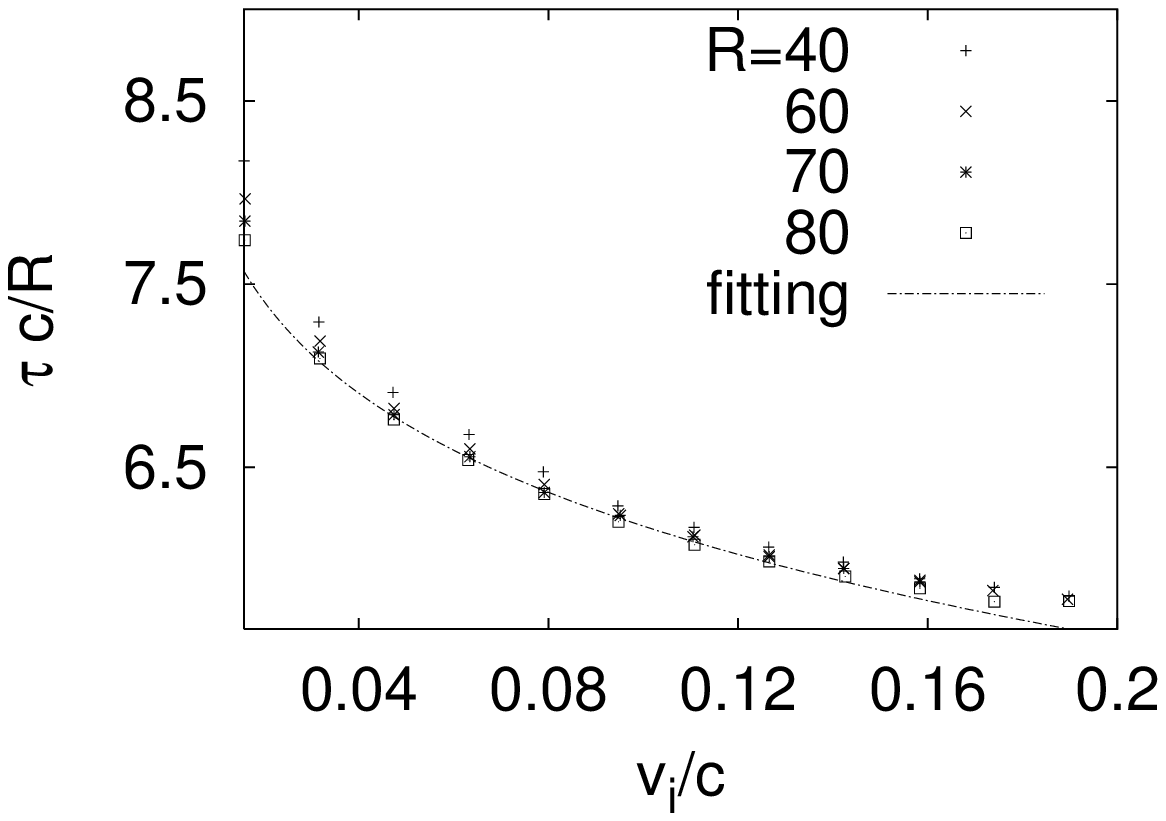}}
 \centerline{\figurename{ 5. authors: Hisao Hayakawa \& Hiroto Kuninaka}}
\label{contact}
\end{figure}

\newpage

\begin{figure}[htbp]
 \epsfxsize=20cm
 \centerline{\epsfbox{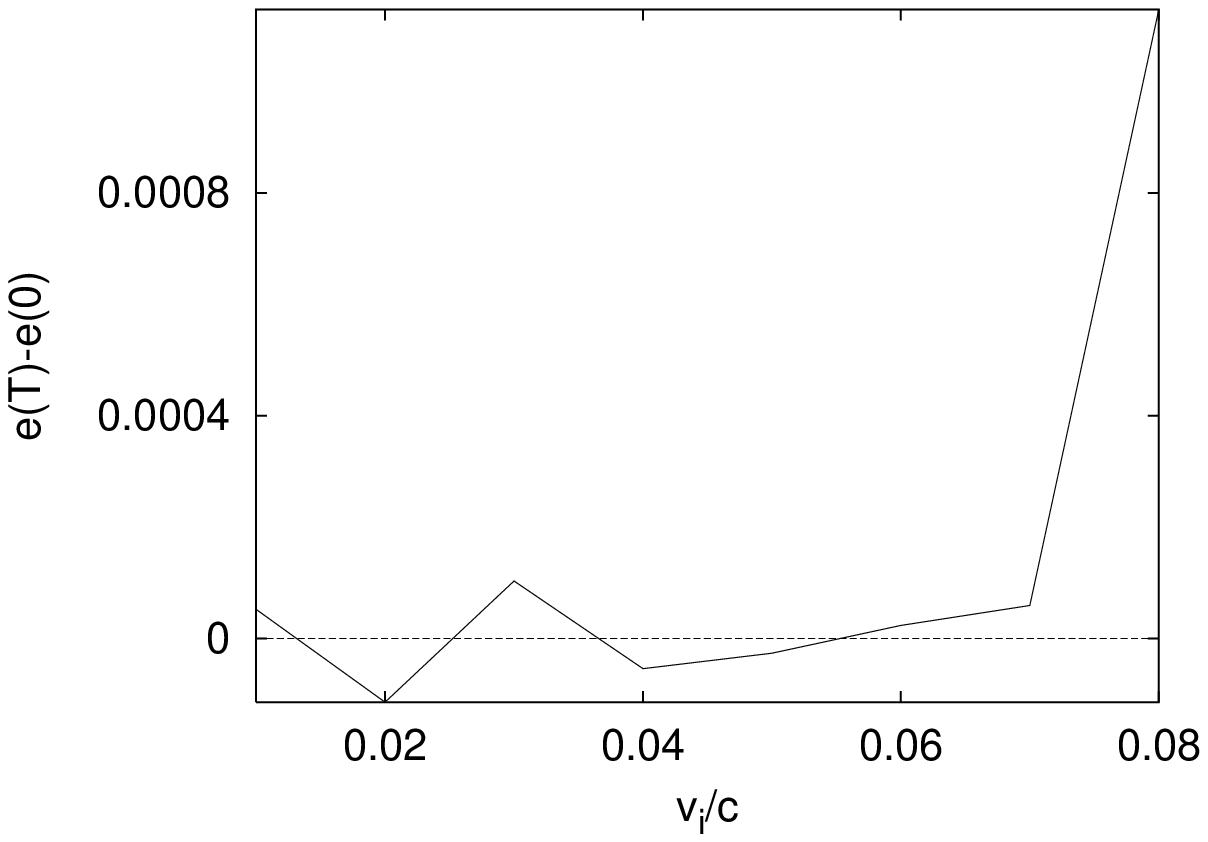}} 
 \centerline{\figurename{ 6. authors: Hisao Hayakawa \& Hiroto Kuninaka}}
 \label{eT-e0}
\end{figure}

\newpage

\begin{figure}[htbp]
 \epsfxsize=20cm
 \centerline{\epsfbox{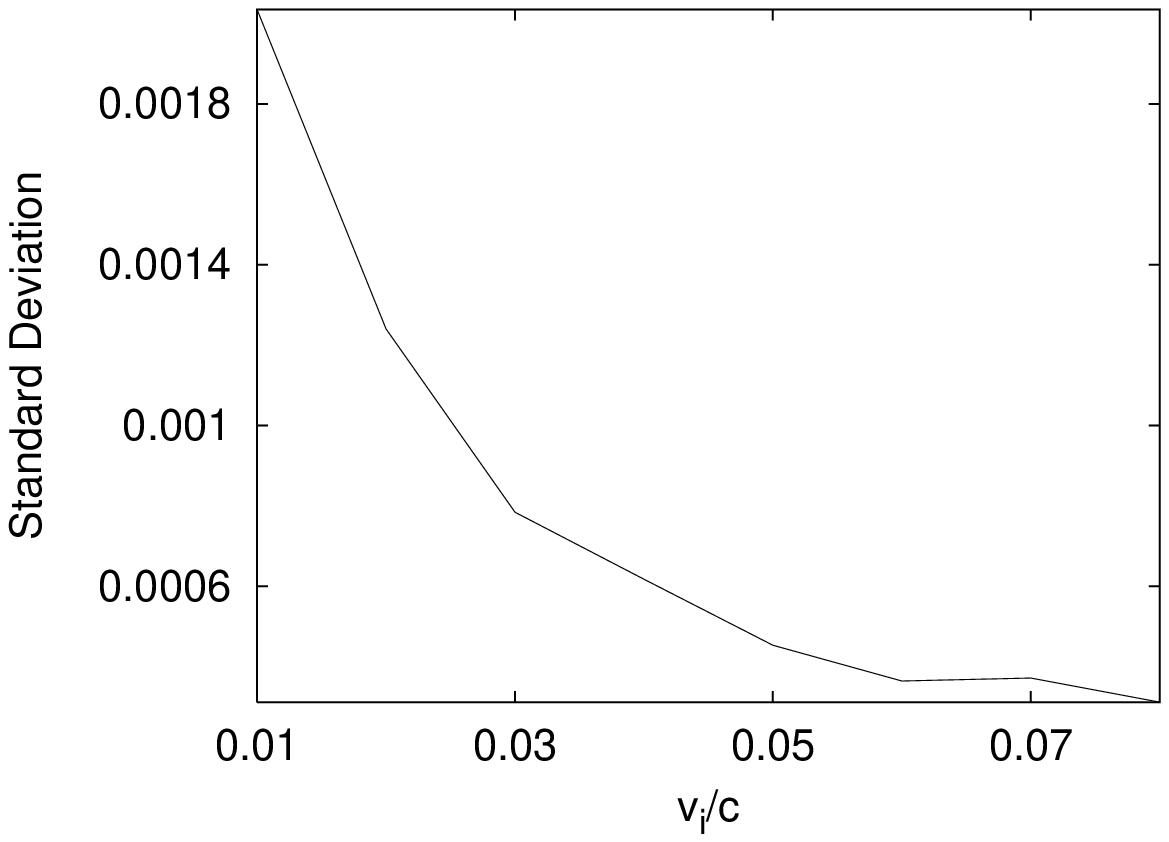}} 
 \centerline{\figurename{ 7. authors: Hisao Hayakawa \& Hiroto Kuninaka}}
 \label{evar}
\end{figure}

\newpage

\begin{figure}[htbp]
 \epsfxsize=20cm
 \centerline{\epsfbox{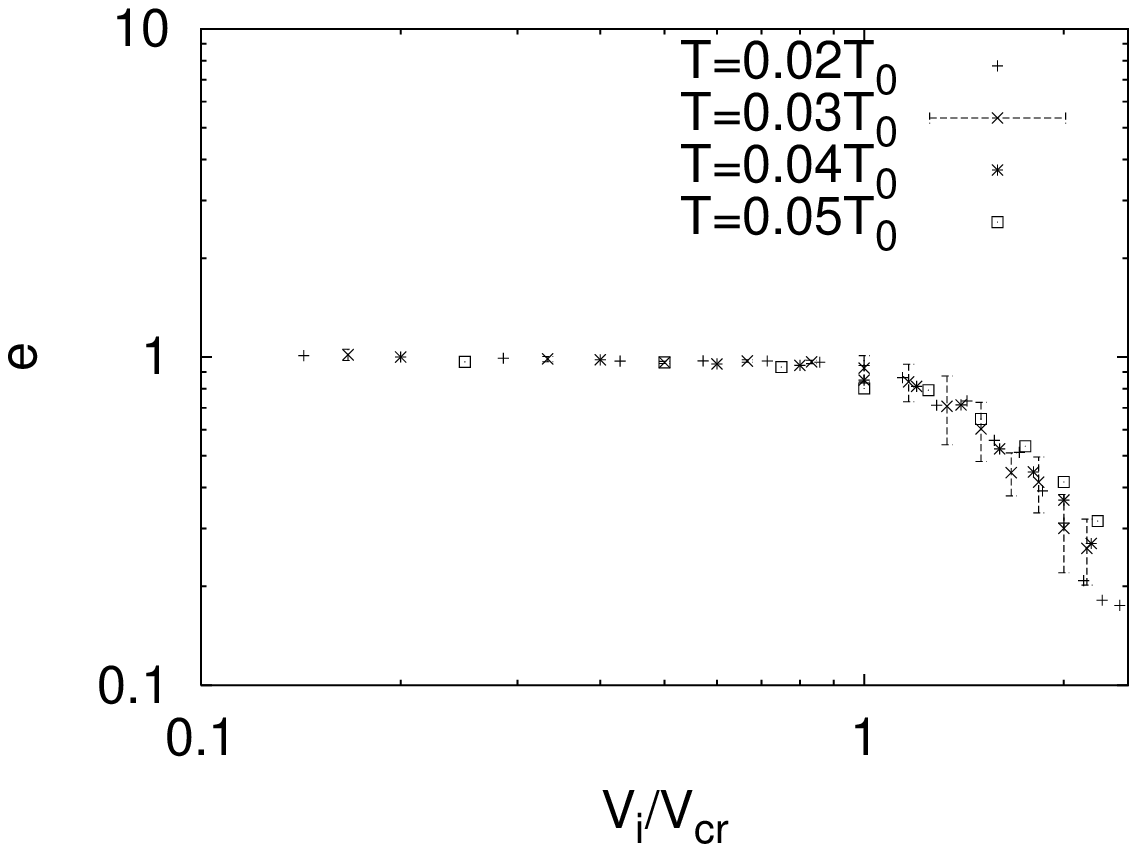}} 
 \centerline{\figurename{ 8. authors: Hisao Hayakawa \& Hiroto Kuninaka}}
 \label{scale2}
\end{figure}

\newpage

\begin{figure}[htbp]
 \epsfxsize=15cm
 \centerline{\epsfbox{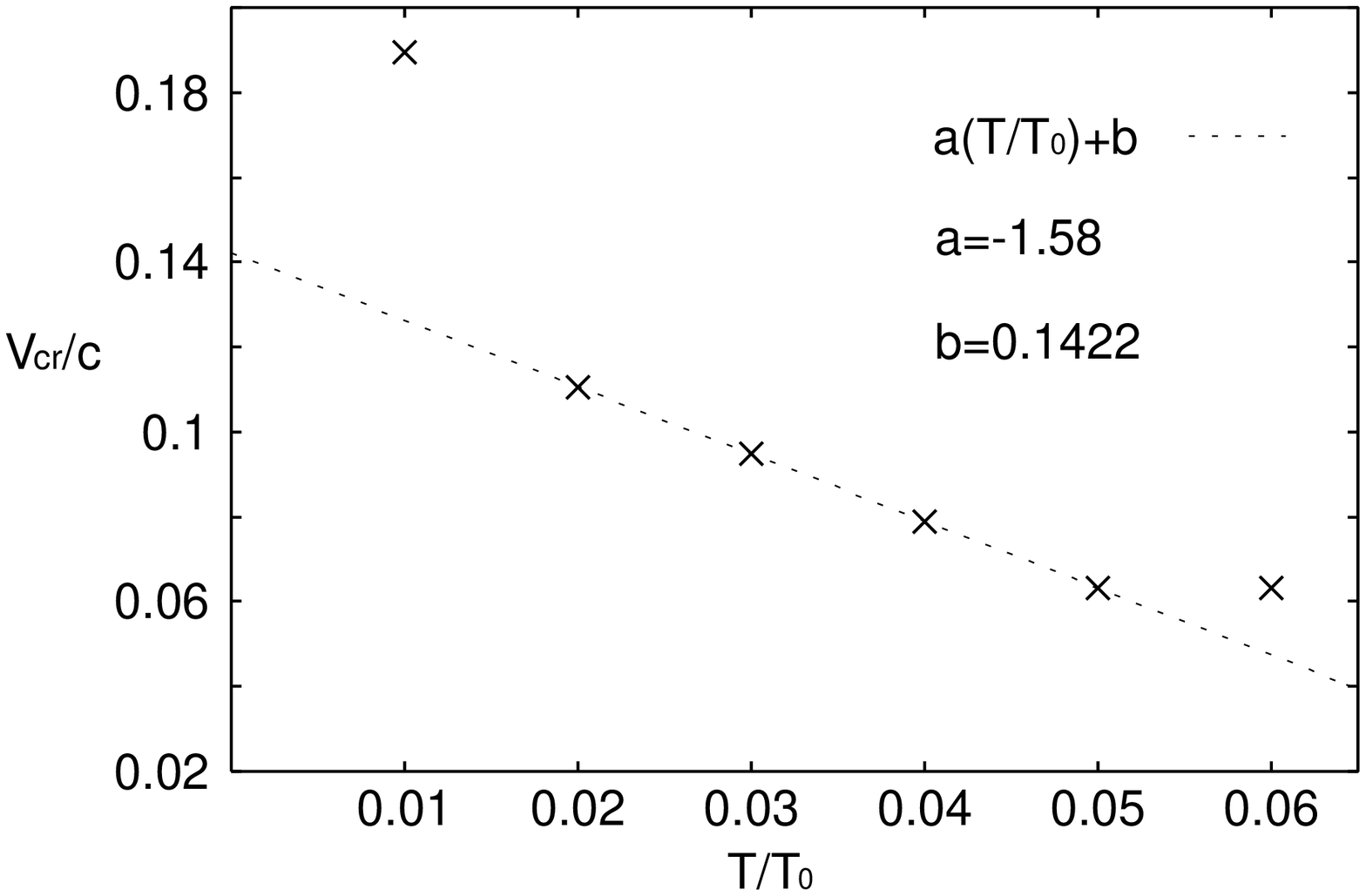}} 
 \centerline{\figurename{ 9. authors: Hisao Hayakawa \& Hiroto Kuninaka}}
 \label{vcr}
\end{figure}

\newpage

\begin{figure}[htbp]
 \epsfxsize=12cm
 \centerline{\epsfbox{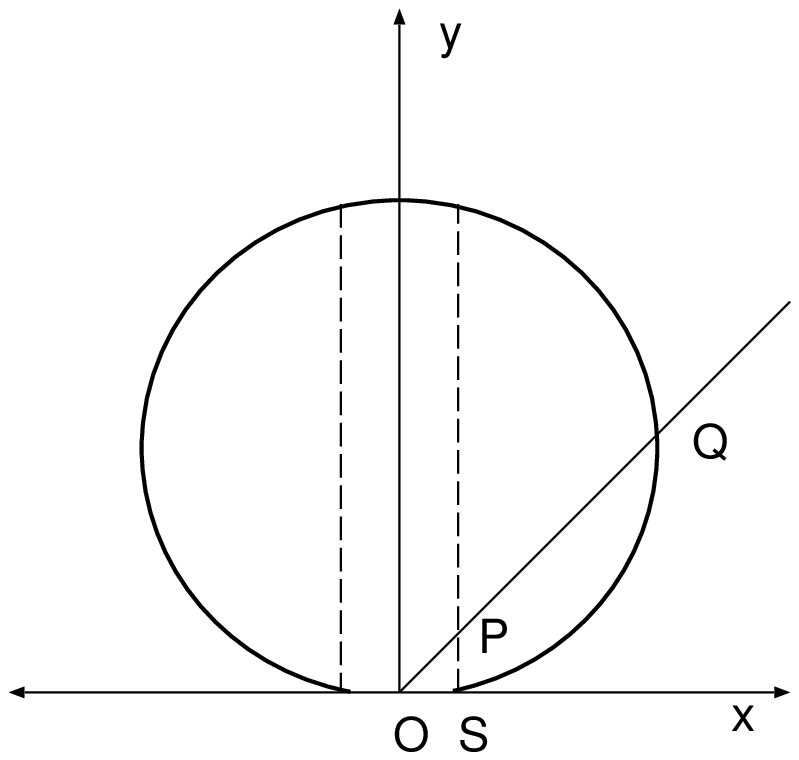}} 
 \centerline{\figurename{ 10. authors: Hisao Hayakawa \& Hiroto Kuninaka}}
 \label{circle-f1}
\end{figure}

\newpage

\begin{figure}[htbp]
 \epsfxsize=15cm
 \centerline{\epsfbox{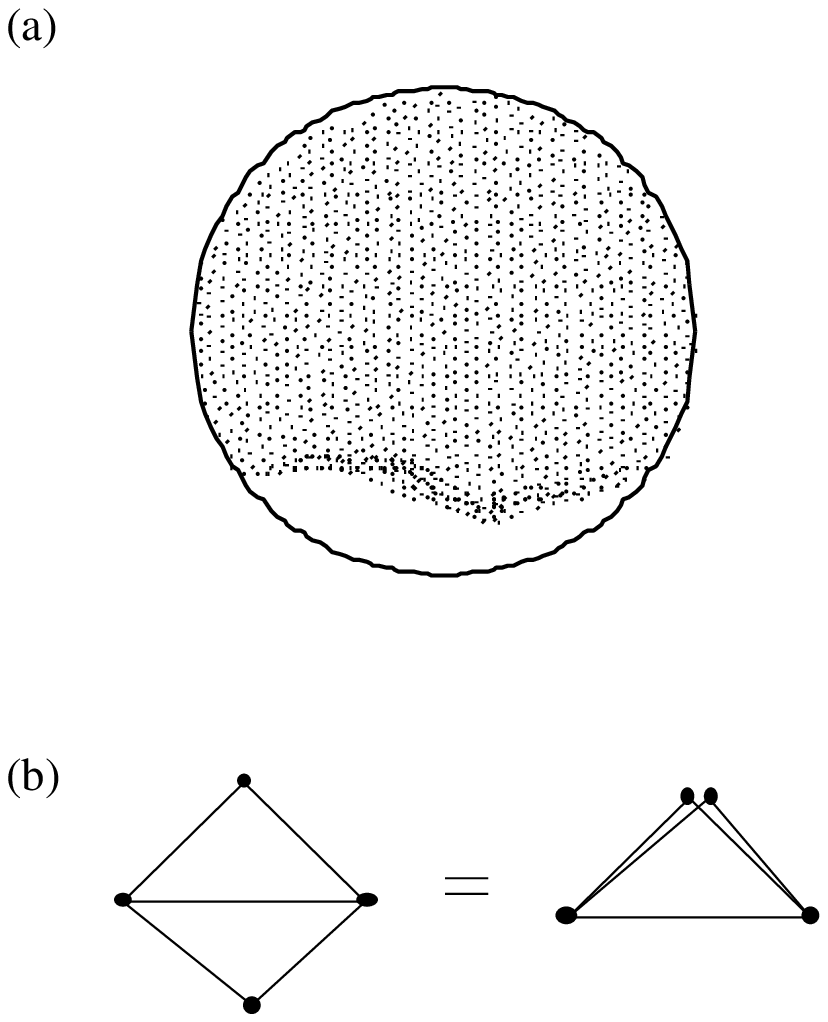}}
 \centerline{\figurename{ 11. authors: Hisao Hayakawa \& Hiroto Kuninaka}}
 \label{deform}
\end{figure}

\end{document}